\documentclass[runningheads]{llncs}
\usepackage{xspace}
\usepackage{amsmath}

\usepackage{xcolor}
\usepackage{mathtools}
\usepackage{float} 
\usepackage{mathpartir}
\usepackage{graphicx}
\usepackage{algpseudocode}
\usepackage[capitalize]{cleveref}
\usepackage{caption}
\usepackage{subcaption}
\usepackage{ulem}
\usepackage{cite}
\usepackage[misc]{ifsym}

\newif\ifExtended\Extendedfalse
\Extendedtrue

\newcommand{\FIXME}[1]{\colorbox{yellow}{\bf FIXME:} \textbf{#1}}

\newcommand{\Powerset}[1]{\mathcal{P}(#1)}
\newcommand{\SD}{\mathsf{SD}}
\newcommand{\defeq}{=}
\newcommand{\DD}[1]{\mathbf{D}({#1})}
\newcommand{\Sig}[2]{\Sigma_{#1}{#2}}
\newcommand{\supp}[1]{\mathsf{supp}({#1})}
\newcommand{\unit}[1]{\mathsf{unit}({#1})}
\newcommand{\ife}[3]{\mathsf{If}\ {#1}\ \mathsf{then}\ {#2}\ \mathsf{else}\ {#3}}
\newcommand{\bind}[2]{\mathsf{bind}({#1},{#2})}
\newcommand{\bindd}{\mathsf{bind}}
\newcommand{\Dproduct}{\otimes}
\newcommand{\pil}{\pi_1}
\newcommand{\pir}{\pi_2}
\newcommand{\padd}[1]{\oplus_{#1}}
\newcommand{\ppadd}{\padd{p}}
\newcommand{\inter}{\cap}
\newcommand{\union}{\cup}

\newcommand{\DV}{\mathbf{DV}}
\newcommand{\RV}{\mathbf{RV}}
\newcommand{\Val}{\mathbf{Val}}
\newcommand{\RanMx}[1]{\mathbf{RanM}({#1})}
\newcommand{\DetM}{\mathbf{DetM}}
\newcommand{\RanM}{\mathbf{RanM}}
\newcommand{\DE}{\mathbf{DE}}
\newcommand{\RE}{\mathbf{RE}}
\newcommand{\sconfig}{\sigma,\mu}

\newcommand{\RC}{\mathbf{RC}}
\newcommand{\CC}{\mathbf{C}}
\newcommand{\SP}{\mathbf{SP}}
\newcommand{\sskip}{\mathbf{skip}}
\newcommand{\assign}[2]{{#1}\gets{#2}}
\newcommand{\rassign}[2]{{#1}\leftarrow_{\$}\mathbf{U}_{#2}}
\newcommand{\ifD}[3]{\mathbf{if}_D\ {#1}\ \mathbf{then}\ {#2}\ \mathbf{else}\ {#3}}
\newcommand{\ifDs}[2]{\mathbf{if}_D\ {#1}\ \mathbf{then}\ {#2}}
\newcommand{\ifR}[3]{\mathbf{if}_R\ {#1}\ \mathbf{then}\ {#2}\ \mathbf{else}\ {#3}}
\newcommand{\ifRs}[2]{\mathbf{if}_R\ {#1}\ \mathbf{then}\ {#2}}
\newcommand{\whileD}[2]{\mathbf{while}_D\ {#1}\ \mathbf{do}\ {#2}}
\newcommand{\whileR}[2]{\mathbf{while}_R\ {#1}\ \mathbf{do}\ {#2}}

\newcommand{\ifNs}[2]{\mathbf{if}\ {#1}\ \mathbf{then}\ {#2}}
\newcommand{\whileN}[2]{\mathbf{while}\ {#1}\ \mathbf{do}\ {#2}}

\newcommand{\llrr}[1]{[[ {#1} ]]}
\newcommand{\semantic}[1]{\llrr{{#1}}}

\newcommand{\semanticstwo}[2]{\semantic{{#1}} ({#2})}
\newcommand{\ssemantics}[1]{\semanticstwo{#1}{\sconfig}}
\newcommand{\Unif}[1]{\mathsf{Unif}_{#1}}
\newcommand{\true}{\mathsf{true}}
\newcommand{\false}{\mathsf{false}}
\newcommand{\op}{\mathsf{op}}
\newcommand{\vset}[1]{\mathsf{vset}({#1})}
\newcommand{\call}[2]{\mathsf{{#1}}({#2})}

\newcommand{\AP}{\mathbf{AP}}

\newcommand{\always}{\top}
\newcommand{\never}{\perp}

\newcommand{\dom}[1]{\mathsf{dom({#1})}}
\newcommand{\sepi}{\mathrel{\rightarrow\!\!\!*}}

\newcommand{\semA}{\models}

\newcommand{\Local}[1]{\mathsf{Ct}({#1})}
\newcommand{\LocalS}{\mathsf{Ct}}
\newcommand{\FV}[1]{\mathsf{FV}({#1})}
\newcommand{\MV}[1]{\mathsf{MV}({#1})}
\newcommand{\ReadV}[1]{\mathsf{RV}({#1})}
\newcommand{\WV}[1]{\mathsf{WV}({#1})}
\newcommand{\imply}{\rightarrow}
\newcommand{\Perm}[1]{\mathsf{Perm}({#1})}
\newcommand{\Count}[1]{\mathsf{Count}({#1})}
\newcommand{\second}[1]{\mathsf{snd}({#1})}

\newcommand{\invs}{\mathsf{inv}}
\newcommand{\eight}{\mathsf{eight}}
\newcommand{\pre}{\mathsf{pre}}
\newcommand{\inv}[1]{\mathsf{inv}({#1})}
\newcommand{\invv}[1]{\mathsf{inv2}({#1})}

\newcommand{\EI}{\mathsf{EI}}

\newcommand{\Uniform}[2]{\mathbf{U}_{#1}[{#2}]}

\newcommand{\triple}[3]{\vdash \{{#1}\}~{#2}~\{{#3}\}}
\newcommand{\NOtriple}[3]{\{{#1}\}~{#2}~\{{#3}\}}

\newcommand{\dassign}[2]{{#1}\leftarrow{#2}}
\newcommand{\ghost}[1]{\textcolor{blue}{#1}}
\newcommand{\assert}[1]{\textcolor{teal}{#1}}
\newcommand{\record}[1]{\ghost{\dassign{Trace}{Trace+#1}}}
\newcommand{\sep}{*}
\newcommand{\size}[1]{\mathsf{size}({#1})}
\newcommand{\Set}[1]{\mathsf{Set}({#1})}
\newcommand{\concat}{\mathrel{+\!+}}
\newcommand{\ReadB}[1]{\mathsf{ReadBucket}({#1})}
\newcommand{\WriteB}[1]{\mathsf{WriteBucket}({#1})}
\newcommand{\find}[1]{\mathsf{find}({#1})}
\newcommand{\Write}{\mathsf{write}}
\newcommand{\select}[1]{\mathsf{select}({#1})}

\newcommand{\linenum}[1]{\textcolor{gray}{\footnotesize{#1}}}

\begin{document}

\title{Combining Classical and Probabilistic Independence Reasoning to Verify the Security of Oblivious Algorithms
\ifExtended (Extended Version) \fi}
\titlerunning{Combining Classical and Probabilistic Independence Reasoning}

\author{Pengbo~Yan\inst{1}(\Letter) \and
Toby~Murray\inst{1} \and
Olga~Ohrimenko\inst{1} \and
Van-Thuan~Pham\inst{1} \and
Robert~Sison\inst{2}}
\authorrunning{P.~Yan et al.}
%
\institute{The University of Melbourne, Melbourne, Australia
\email{pengboy@student.unimelb.edu.au, \{toby.murray,oohrimenko,thuan.pham\}@unimelb.edu.au} \and
UNSW Sydney, Sydney, Australia, 
\email{r.sison@unsw.edu.au}}
\maketitle


\begin{abstract}
  We consider the problem of how to verify the
  security of probabilistic oblivious algorithms formally and systematically. Unfortunately,
  prior program logics fail to support a number of complexities that feature in the semantics and invariant
  needed to verify the security of many practical probabilistic
  oblivious algorithms. We propose
  an approach based on reasoning over perfectly oblivious
  approximations, using a program logic that combines both
  classical Hoare logic reasoning and probabilistic
  independence reasoning to support all the needed features.
  We formalise and prove our new logic
  sound in Isabelle/HOL and apply our approach to
  formally verify the security of
  several challenging case studies beyond the reach of
  prior methods for proving obliviousness.
\end{abstract}


\section{Introduction}\label{sec:intro}

Side-channel attacks allow attackers to infer sensitive information 
by eavesdropping on a program's execution, 
when the sensitive data are not directly observable (e.g.\ because they are encrypted). For example, sensitive documents or secret images can be
reconstructed by only observing a program's memory access pattern \cite{Leakage, SGX, Last-Level}. 
Many algorithms are charged with the protection of secrets
in application contexts where such attacks are realistic,
for example, cloud computing\cite{OlyaSampling,Opaque}, secure processors \cite{rawORAM, PHANTOM} and multiparty computation \cite{ObliVM}.

The goal of an \emph{oblivious algorithm}
(e.g. path ORAM\cite{pathORAM}, Melbourne shuffle \cite{Melbourne}) 
is to hide its secrets from an attacker that can observe memory accesses. \emph{Probabilistic} oblivious algorithms aim to do so while achieving better performance than \emph{deterministic} oblivious algorithms. The various programming disciplines to defend against such attacks for deterministic 
algorithms~\cite{PC-Security,ConstantTime} often lead to poor performance: e.g.\ to hide the fact
that an array is accessed at a certain position, one
may have to iterate over the entire array~\cite{FaCT}. Probabilistic oblivious algorithms avoid this inefficiency by performing
random choices at runtime to hide
their secrets from attackers more efficiently. Unfortunately, probabilistic methods for achieving obliviousness are error prone and some have been shown insecure, as a result requiring non-trivial fixes~\cite{10.5555/2095116.2095129,10.1007/978-3-642-22012-8_46}.

In this paper we develop a program logic to verify the
security of probabilistic oblivious algorithms formally and systematically.
We adopt the standard threat model for such programs,
in which the attacker is assumed to be able to infer the memory
access pattern (e.g.~either by explicitly observing memory requests in case of 
untrusted/compromised operating system or by measuring the time its own memory
accesses take due to shared resources like caches)~\cite{SquareRoot,pathORAM,Stash,191010}.

Although some previous works \cite{PSL, Oblicheck, OADT, RObliv} exist, many oblivious algorithms 
have complex semantics and invariants that are beyond the reach of those prior methods to reason about.
For example, path ORAM \cite{pathORAM} maintains an invariant stating that virtual addresses are independent of 
each other and of the program's memory access patterns; whereas
the oblivious sampling algorithm \cite{OlyaSampling} contains secret- or random-variable-dependent random choices, conditional branches and loops,
whose details we introduce in \cref{sec:overview} and \cref{sec:examples}.

Also, to achieve efficiency, some oblivious algorithms \cite{pathORAM,pathOheap,Melbourne} forgo perfection and
have a very small probability of failure, which means
that they do not perfectly hide their secrets. Fortunately,
they are intentionally designed so that the failure probability is bounded
by some negligible factor (e.g.\ of the size of the secret data), meaning that they are secure in practice.
Following prior work \cite{pathORAM,Melbourne}, this means that we can prove them secure by reasoning over
\emph{perfectly oblivious approximations}, the theoretical and perfect version of the practical algorithms that are free of failure by construction
(\cref{sec:security} justifies this claim).
Proving negligible error probability bounds on oblivious algorithms is an important goal, but is out of scope of this present
work.

Reasoning over the perfectly oblivious approximations requires an approach that
supports for all of the following:
\begin{itemize}
	\item Assertions that describe probability distributions and independence;
	\item Reasoning about dynamic random choices over secrets and random variables -- e.g. a random choice of integers from 1 to random secret variable $s$; 
	\item Reasoning about branches that depend on secret random variables;
	\item Reasoning over loops that have a random  number of iterations.
\end{itemize}
Our approach that addresses these challenges simultaneously.

Following preliminaries (\cref{sec:pre}), in~\cref{sec:logic} we build a program logic
that combines \emph{classical}
and \emph{probabilistic} reasoning to address the aforementioned challenges, which we
prove sound in Isabelle/HOL. Our
logic is situated atop the Probabilistic Separation Logic (PSL)~\cite{PSL}; proving the soundness of our logic
revealed several
oversights in PSL~\cite{PSL}, which we fixed (see \cref{unsound}). 

To our knowledge, the reasoning our logic supports is beyond all prior methods
for verifying obliviousness, including PSL~\cite{PSL}, ObliCheck~\cite{Oblicheck}, $\lambda_{\mathit{OADT}}$~\cite{OADT}, and $\lambda_{\mathit{obliv}}$~\cite{RObliv}.
The combination of classical
and probabilistic reasoning also makes our logic more expressive than previous
probabilistic Hoare logics (e.g., \cite{pHL_Den}, VPHL~\cite{VPHL} and pRHL~\cite{Easycrypt})
which, because they lack assertions for describing distributions and independence, are
ill-suited to direct proofs of obliviousness.

Finally, we demonstrate the power of our
logic by applying it on pen-and-paper to verify, for the first time, the obliviousness of several non-trivial case studies (\cref{sec:examples}). Their verification is a significant achievement in that they
constitute the fundamental
building blocks for secure oblivious systems.

\section{Overview} \label{sec:overview}

\subsection{Challenges for verification} \label{sec:motivation}

Many probabilistic oblivious algorithms use probabilistic independence as a core intermediate condition to prove their obliviousness informally on pen and paper \cite{pathORAM, Melbourne, pathOheap, OlyaSampling}, which is intuitive and simple. However, such algorithms present a range of challenges for verifying their obliviousness formally and systematically.

\begin{figure}[t]
		\[
		\begin{array}{@{}r@{\ \ \ \ \ \ }l}
			&\hspace*{-0.4cm}\assert{\text{Let } \eight(i) = \{[x_0, x_1, \cdots , x_{i-1}]~|~\forall j. ~~0\le x_j \le 7\} }\\
			&\hspace*{-0.4cm}\assert{\text{Let } \pre = \{ \forall i\in \{0,1,\cdots,n-1\}.~~S[i]\in \{0,1\} \} }\\
			&\hspace*{-0.4cm}\assert{\text{Let } \invs(x) = \{ \Local{\pre\land i\le n} \land \Uniform{\eight(x)}{O} \} }\\
			&\hspace*{-0.4cm}\mathsf{synthetic}(S,O,n):\\
			&\assert{\{\Local{\pre \land O = []}\}}\\
			\linenum{1}&\rassign{A[0]}{\{0,1,2,\cdots,7\}};\\
			&\assert{\{\Local{\pre \land O = []} \land \Uniform{\{0\cdots7\}}{A[0]}\}}\\
			\linenum{2}&\rassign{A[1]}{\{0,1,2,\cdots,7\}};~\dassign{i}{0};\\
			&\assert{\{\Local{\pre \land O = [] \land i = 0} \land \Uniform{\{0\cdots 7\}}{A[0]} \sep \Uniform{\{0\cdots 7\}}{A[1]}\}}\\
			\linenum{3}&\whileN{i<n}{}\hfill\assert{\text{because $\eight$(0) = \{[]\}}}\\
			&\hspace*{0.4cm} \assert{\{\invs(i)\sep \Uniform{\{0\cdots 7\}}{A[S[i]]}\sep\Uniform{\{0\cdots 7\}}{A[1-S[i]]}\}}\\
			\linenum{4}&\hspace*{0.4cm} \dassign{O}{O + A[S[i]]};\hfill\assert{\text{using proposition 1.8}}\\
			&\hspace*{0.4cm} \assert{\{\invs(i+1)\sep\Uniform{\{0\cdots 7\}}{A[1-S[i]]}\}}\\
			\linenum{5}&\hspace*{0.4cm} \dassign{m}{8};~\dassign{j}{0};\\
			&\hspace*{0.4cm} \assert{\{\invs(i+1)\sep\Uniform{\{0\cdots 7\}}{A[1-S[i]]} \land \Local{m=8\land j=0}\}}\\
			\linenum{6}&\hspace*{0.4cm} \whileN{A[S[i]]>j}{}\\
			&\hspace*{0.8cm} \assert{\{\Local{m > 7\land m~\%~8 = 0}\}}\\
			\linenum{7}&\hspace*{0.8cm} \dassign{m}{m*2};~\dassign{j}{j+1};\ \\
			\linenum{8}&\hspace*{0.8cm} \ifNs{(j+S[i])~\%~3 == 0}{}\ \\
			\linenum{9}&\hspace*{1.2cm} \dassign{j}{j+1};\ \\
			&\hspace*{0.4cm} \assert{\{\Local{m > 7\land m~\%~8 = 0}\}}\hfill\assert{\text{using Const rule around the loop}}\\
			&\hspace*{0.4cm} \assert{\{\invs(i+1)\sep\Uniform{\{0\cdots 7\}}{A[1-S[i]]}\land\Local{m > 7\land m~\%~8 = 0}\}}\\
			\linenum{10}&\hspace*{0.4cm} \rassign{t}{\{1,2,3,\cdots,m\}};\hfill\assert{\text{using RSample}}\\
			&\hspace*{0.4cm} \assert{\{\invs(i+1)\sep\Uniform{\{0\cdots 7\}}{A[1-S[i]]}\land\Uniform{\{0\cdots 7\}}{t~\%~8}\}}\\
			\linenum{11}&\hspace*{0.4cm} \dassign{A[S[i]]}{t~\%~8};\hfill\assert{\text{using Rassign, Unif-Idp rule}}\\
			&\hspace*{0.4cm} \assert{\{\invs(i+1)\sep\Uniform{\{0\cdots 7\}}{A[1-S[i]]}\sep\Uniform{\{0\cdots 7\}}{A[S[i]]}\}}\\
			\linenum{12}&\hspace*{0.4cm} \dassign{i}{i+1};\ \\
			&\assert{\{\invs(n)\}}
		\end{array}
		\]
		\caption{\label{fig:motivation}Verification of the motivating algorithm.}
\end{figure}

We have constructed the example algorithm in \cref{fig:motivation} to illustrate in a simplified form the kinds of complexities that will feature in the semantics and invariants needed to prove our case studies (\cref{sec:examples}). The teal-coloured parts show the verification and will be introduced in the next subsection.
Our synthetic algorithm takes an input array $S$ with size $n$ containing secret elements: each either 0 or 1. The list $O$ is empty initially but will be filled with some data later. We want to prove $O$ will not leak any information about $S$. The synthetic algorithm first initialises an array $A$ with two random values sampled from the integers between 0 and 7, then it has a nested loop showing the following challenges:

\begin{enumerate}
\item The outer loop iterates $n$ times where the $i$th iteration will append $A[S[i]]$ to $O$ (line $\linenum{4}$). It simulates a simplified version of path ORAM \cite{pathORAM}, which maintains an invariant that virtual addresses are independent of each other and of the program's memory access patterns. The secret $S$ can be seen as a sequence of secret virtual addresses and the output $O$ represents the memory access pattern. We need to prove an invariant that the elements in $O$ are independent of each other and independent of each element~$A[S[i]]$ appended to $O$ by the outer loop. Note: the assignment on
  line~$\linenum{4}$ breaks the independence between $O$ and~$A[S[i]]$, so
  lines $\linenum{4}$--$\linenum{11}$ update $A[S[i]]$ with a fresh random value
  to re-establish the independence for the next loop iteration. This
  ensures $O$ is independent of $S$ and will not leak secret information.
	
\item After initialising $m$ with 8 on line $\linenum{5}$, we have the inner loop containing a \emph{probabilistic and secret-dependent if-conditional}. Its secret dependence makes the control flow different over different values of the secret. The iteration count for
  the inner loop is \emph{truly random}, depending on $A[S[i]]$ (where each iteration doubles $m$ and increases $j$ by 1 or 2 depending on whether $j + S[i] \% 3 = 0$). These kinds of loops and conditionals are common in real-world oblivious algorithms (\cref{sec:examples}), yet necessarily complicate reasoning.
	
	\item On line $\linenum{10}$,
	  the algorithm makes what we call a \emph{dynamic random choice}, which is one over a truly random set (here, from 1 to the random variable $m$), assigning the chosen value to~$t$. Then, (line~$\linenum{11}$) $A[S[i]]$ is assigned $t~\%~8$. This requires 
          reasoning that
          $t~\%~8$ satisfies the uniform distribution on $\{0\cdots 7\}$, because $m$ is certainly a multiple of 8. Dynamic random choices are also common in real-world oblivious algorithms, as \cref{sec:examples} demonstrates.
\end{enumerate}

Lines $\linenum{5-11}$ are derived from the oblivious sampling algorithm~\cite{OlyaSampling} (\ifExtended
\cref{sec:OlyaSampling})
\else
Appendix C.2 of the extended version of the paper on Arxiv)
\fi to demonstrate challenges 2 and 3.

\subsection{Mixing Probabilistic and Classical Reasoning}\label{sec:nondet}

We show how to construct a program logic that combines classical and
probabilistic (and independence) reasoning over different parts of
the program so that it can verify our running example, as shown in \cref{fig:motivation}.
Namely, certain parts of the algorithm (lines $\linenum{1,2,4,10}$) require careful
probabilistic reasoning, while others
do not, but that each style of reasoning can benefit the other.

Our program logic is constructed by situating these ideas in
the context of the Probabilistic Separation Logic (PSL)~\cite{PSL}. PSL is an existing program logic for reasoning about
probabilistic programs. PSL employs the separating conjunction (here written $\star$)
familiar from separation logic~\cite{EarlySL} to capture when two
probability distributions are independent. In situating our
work atop PSL we extend its assertion forms with the new $\Local{\cdot}$ assertion, to
capture classical information. More importantly, however, we
significantly extend the resulting logic with a range of novel reasoning principles
for mixing classical and probabilistic reasoning embodied in a suite of new rules
(\cref{rules}), which we will present more fully in \cref{sec:logic}. These
new rules show how classical reasoning (captured by~$\Local{\cdot}$ assertions)
can be effectively harnessed, and allow reasoning about dynamic random choices,
secret-dependent if-statements, and random loops, making our logic significantly
more applicable than PSL; while leveraging PSL's support for intuitive reasoning about probability distributions makes our logic
also more expressive than prior probabilistic program logics \cite{Easycrypt, pHL_Den, VPHL}.
We also harness the close interaction between classical and
probabilistic reasoning to allow
new ways to prove security (e.g., the \textsc{Unif-Idp} rule and the final
proposition of \cref{prop}, which will be introduced in \cref{rules} and \cref{sec:assertions}), and new ways to reason about random sampling
(embodied in the \textsc{RSample} rule, \cref{rules}). Each represents a non-trivial insight, and
all are necessary for reasoning about real-world oblivious algorithms
(\cref{sec:examples}). 
The increase in expressiveness, beyond prior probabilistic
program logics~\cite{PSL,pHL_Den, VPHL, Easycrypt}, within a principled and clean
extension of PSL attests to the careful design of our logic.

The combination of classical and probabilistic
reasoning means that our logic tracks two kinds of
\emph{atomic assertions}, as follows.

\textbf{Certain Assertions.} Classical reasoning is supported
by certain assertions $\Local{e_r}$ that state that some property~$e_r$ (which may mention random variables) is true with
absolute certainty, i.e.\ is true in all memories supported by
the current probabilistic state of the program. With certain assertions and classical reasoning, our logic can reason about
\textbf{loops with random iteration numbers and randomly secret-dependent if statements}. Doing so requires distinguishing classical
from distribution (independence) assertions, because the latter are ill-suited for reasoning about random loops and conditionals.

For example, from line $\linenum{5}$ to $\linenum{9}$, although the random loop and the probabilistic- and secret-dependent if statement complicate the algorithm, we only need classical reasoning to conclude that after the loop $m$ is certainly a multiple of 8 (using the \textsc{RLoop} and \textsc{RCond} rules in \cref{rules}, which have the classic form). This information is sufficient to verify the remainder of the algorithm.

\textbf{Distribution Assertions.} On the other hand, reasoning about probability distributions
is supported by distribution assertions, which we adopt and extend from PSL:
for a set expression~$e_d$ (which is allowed to mention
non-random program variables), $\Uniform{e_d}{e_r}$
states that expression~$e_r$ is uniformly distributed over
the set denoted by~$e_d$ in the sense that when $e_r$ is
evaluated in the current probabilistic state of the program
it yields a uniform distribution over the evaluation of
$e_d$. We define these concepts formally later in \cref{sec:assertions} (see \cref{defn:atomic-assertions}). With this reasoning style, we support \textbf{dynamic random choice} (e.g. line $\linenum{10}$, the value is sampled from a truly probabilistic set), which is not supported by previous works \cite{PSL, Oblicheck, OADT, RObliv, Easycrypt, pHL_Den, VPHL}. Note that we require $e_d$ to be deterministic here because if $e_d$ can be probabilistic, then it means a probabilistic expression satisfies a \emph{uniform distribution on a probabilistic set}---a clear contradiction.

For example, at line $\linenum{10}$, even if we do not specify the detailed distribution of $m$, we can conclude $t~\%~8$ satisfies the uniform distribution on the set $\{0\cdots7\}$, as $m$ is certainly a multiple of 8, by an argument based on our concept of an \emph{even partition} (\cref{defn:even-partition}). This reasoning is supported by our novel \textsc{RSample} rule (\cref{rules}). Here, it requires that all the possible sets (in this case, $\{1\cdots 8\}$ or $\{1\cdots 16\}$ or ...) over which~$t$ was sampled, can each be evenly mapped to (and thus partitioned by) the target set (here $\{0\cdots 7\}$) by the applied function (here $\% 8$). Thus $t~\%~8$ must satisfy the uniform distribution on $\{0\cdots 7\}$. 

\paragraph{Unifying classical and probabilistic independence reasoning} Another important
feature of our logic is that it allows independence to be \emph{derived} by leveraging
classical reasoning. For example, considering line $\linenum{10, 11}$, if a variable ($A[S[i]]$) always satisfies the same distribution (uniform distribution on $\{0\cdots 7\}$) over any possible values of some other variables ($O$ and $A[1-S[i]]$), then the former is independent of the latter (because $O$ and $A[1-S[i]]$ will not influence the values of $A[S[i]]$).
The new rule \textsc{Unif-Idp} (\cref{rules}) embodies this reasoning (where
$\star$ denotes independence and $\DD{}$ stands for an arbitrary distribution).\footnote{In this case we cannot use PSL's  frame rule because $m$ is not independent of $A$.}

Our logic also includes a set of useful propositions
(\cref{prop})
that aid deriving independence information from classical reasoning.

Returning to the example,
with the conclusion that $A[S[i]]$ is independent of other variables, we can construct the loop invariant of the outer loop ($\invs(i)$) stating that the output array $O$ always satisfies a uniform distribution following the $i$th iteration, which is captured by $\eight(i)$. We
use the final proposition of \cref{prop} here. Intuitively, this proposition says given a reversible function (whose inputs can be decided by looking at its outputs, e.g. array appending), if its two inputs satisfy uniform distribution and are independent of each other, then the result of the function should satisfy the uniform distribution on the product (by the function) of the two inputs' distribution.

By the invariant, we can conclude finally the output array always satisfies the uniform distribution on $\eight(n)$, regardless of secret $S$, which means the output will not leak any secret information.

\section{Preliminaries}
\label{sec:pre}

\subsection{Programming Language and Semantics}
In this paper we define a probability distribution over a countable set $A$ is a function $\mu:A\rightarrow[0,1]$ where $ \Sig{a\in A}{\mu(a)} = 1$. We write $\mu(B)$ for $\Sig{b\in B}{\mu(b)}$ where $B$ can be any subset of $A$ and $\DD{A}$ for the set of all distributions over $A$.\par

The \emph{support} of a distribution~$\mu$, $\supp{\mu}$, is the set of all elements whose probability is greater than zero, $\{a\in A\ |\ \mu(a)>0\}$.\par

A unit distribution over a single element, $\unit{a}$, is $(\lambda x.\ \ife{a=x}{1}{0})$. A uniform distribution over a set, $\Unif{S}$, is $(\lambda x.\ \ife{x\in S}{1/|S|}{0})$.\par

Given a distribution $\mu$ over $A$ and a function $f$ from elements of $A$ to a distribution, $f:A\rightarrow \DD{B}$, we define $\bind{\mu}{f}\defeq\lambda b.\ \Sig{a\in A}{\mu(a)\cdot f(a)(b)}$, used to give semantics to random selections and assignments to random variables.\par

Given two distributions $\mu_A$ and $\mu_B$ over the sets $A$ and $B$, we define $\mu_A \Dproduct \mu_B \defeq \lambda a, b.\ \mu_A(a)\cdot \mu_B(b)$. Given a distribution $\mu$ over $A\times B$, we define $\pil(\mu) \defeq \lambda a.\ \Sig{b\in B}{\mu(a,b)}$ and $\pir(\mu) \defeq \lambda b.\ \Sig{a\in A}{\mu(a,b)}$. We say these two distributions are \emph{independent} if and only if $\mu = \pil(\mu) \Dproduct \pir(\mu)$.\par

Given a distribution $\mu$ over some set $A$, and $S\subseteq A$ where $\mu(S) > 0$, let $E\subseteq A$, we define $(\mu | S) \defeq \lambda E.\ \frac{\mu(E\cap S)}{\mu(S)}$, used to give semantics to conditional statements, as is the following. Given two distribution $\mu_1, \mu_2$, and a number $p\in [0,1]$, we define $\mu_1\ppadd \mu_2\defeq \lambda x.\ p\cdot\mu_1(x) + (1-p)\cdot\mu_2(x)$. When $p$ is 1 or 0, we unconditionally define the result to be $\mu_1$ or $\mu_2$ respectively.

Same as PSL's memory
model, we also distinguish \emph{deterministic} from
\emph{random} variables: only the latter can be
influenced by random selections (i.e.\ by probabilistic
choices).
We define $\DV$ as a countable set of deterministic variables and $\RV$ as a countable set of random variables, disjoint from $\DV$.

Let $\Val$ be the countable set of values, which we assume
contains at least the values $\true$ and $\false$.
When applying our logic, we will freely assume it
contains integers, lists, sets, and any other
standard data types as required.
Let $\op$ be a set of operations on values, including binary functions on values of type $(\Val \times \Val) \rightarrow \Val$. In practice, we will assume it includes the
standard arithmetic, list and set operations, and others as required. Finally, let $\vset{}$ be a function of type $\Val \rightarrow \Powerset{\Val}$, taking one value and returning a non-empty, finite set of values, for giving
semantics to dynamic random choice. \label{vset}

Then let $\DetM \defeq \DV\rightarrow \Val$ be the set of
deterministic memories, and $\RanM \defeq \RV\rightarrow \Val$ the set of random variable memories. A semantic configuration
is a pair $(\sigma, \mu)$, where $\sigma \in \DetM$ and $\mu\in\DD{\RanM}$ (a probability distribution over $\RanM$). Configurations represent
program states. 

As with program variables, we define sets of
deterministic and random expressions, denoted $\DE$ and
$\RE$ respectively. $\DE$ cannot mention random variables.

\begin{definition}[Expressions]
	Expressions are either deterministic or random, defined as follows:\\
	\centerline{Deterministic expressions: $\DE\ni e_d ::= \Val\ |\ \DV\ |\ \op\ \DE\ \DE$}
	\centerline{Random expressions: $\RE\ni e_r ::= \Val\ |\ \DV\ |\ \RV\ |\ \op\ \RE\ \RE$}
\end{definition}

Note that $\DE$ is a subset of $\RE$. Given a deterministic memory $\sigma$ and a random variable memory $m$, we write $\semantic{e_r}(\sigma,m)$ as the evaluation of expression $e_r$. Expression
evaluation is entirely standard and its definition is omitted
for brevity.
The evaluation of deterministic expressions~$e_d$
depends only on the deterministic memory~$\sigma$ and so we often
abbreviate it $\semantic{e_d}\sigma$.

Following the distinction between deterministic and random variables,
the programming language also
distinguishes deterministic and random conditionals and loops. We define two sets of program commands for our language, where $\CC$ is the complete set of commands and $\RC$ is a subset of $\CC$ containing so-called ``random''
commands that cannot assign to
deterministic variables.
We write $\ifDs{b}{c}$ to abbreviate $\ifD{b}{c}{\sskip}$ and likewise for $\ifRs{b}{c}$.
As with PSL, our logic is defined for programs that
always terminate.

\noindent\begin{tabular}{l@{}l}
	\begin{minipage}{0.5\textwidth}
		$\RC\ni c::= \sskip\ |\ \assign{\RV}{\RE}$\\
		\hspace*{0.5cm}$|\ \rassign{\RV}{\RE}\ |\ \RC;\RC\ $\\
		\hspace*{0.5cm}$|\ \ifD{\DE}{\RC}{\RC}$\\
		\hspace*{0.5cm}$|\ \ifR{\RE}{\RC}{\RC} $\\
		\hspace*{0.5cm}$|\ \whileD{\DE}{\RC}$\\
		\hspace*{0.5cm}$|\ \whileR{\RE}{\RC} $\\
	\end{minipage} &
	\begin{minipage}{0.5\textwidth}
		$\CC\ni c::= \sskip\ |\ \assign{\DV}{\DE}\ $\\
		\hspace*{0.5cm}$|\ \assign{\RV}{\RE}\ |\ \rassign{\RV}{\RE}\ |\ \CC;\CC\ $\\
		\hspace*{0.5cm}$|\ \ifD{\DE}{\CC}{\CC}$\\
		\hspace*{0.5cm}$|\ \ifR{\RE}{\RC}{\RC} $\\
		\hspace*{0.5cm}$|\ \whileD{\DE}{\CC} $\\
		\hspace*{0.5cm}$|\ \whileR{\RE}{\RC} $\\
	\end{minipage}
\end{tabular} 

In practical verification, given an algorithm, we try to set all the variables as deterministic variables at the beginning. Then, all the variables sampled from the uniform distribution or assigned by an expression containing random variables must be random variables. All the loop and IF conditions containing random variables must be random loop/IF. All the variables assigned in a random loop/IF must be random variables. We repeat the above process until no variable and loop/IF statements will change their type.

The semantics (\cref{semantic}) of a command~$c \in \CC$ is denoted
$\semantic{c}$, which is a configuration transformer
of type $(\DetM\times\DD{\RanM})\rightarrow (\DetM\times\DD{\RanM})$.
Our programming language extends that of PSL by allowing
dynamic random choice, in which a value is chosen from a set
denoted by an random
expression~$e_r \in \RE$ rather than a constant set.
We also add random loops, whose condition can depend on random expressions (rather
than only deterministic expressions as in PSL).
These improvements increase the expressivity of the language,
necessary to capture the kinds of practical oblivious
algorithms that we target in \cref{sec:examples}.
\begin{figure}
	\begin{align*}
		\ssemantics{\sskip}&\defeq(\sconfig)\\
		\ssemantics{\assign{x_d}{e_d}}&\defeq(\sigma[x_d\mapsto \semantic{e_d}\sigma], \mu)\\
		\ssemantics{\assign{x_r}{e_r}}&\defeq
		(\sigma, \bindd(\mu,m\mapsto \unit{m[x_r\mapsto \semantic{e_r}(\sigma,m)]}))\\
		\ssemantics{\rassign{x_r}{e_r}}&\defeq
		(\sigma, \bindd(\mu,m\mapsto\bindd(\Unif{\vset{\semantic{e_r}(\sigma,m)}},u\mapsto \unit{m[x_r\mapsto u]})))\\
		\ssemantics{c;\ c'}&\defeq\semanticstwo{c'}{\ssemantics{c}}\\
		\ssemantics{\ifD{b}{c}{c'}}&\defeq
		\begin{cases}
			\ssemantics{c} & :\semantic{b}\sigma \neq \false \\    
			\ssemantics{c'} & :\semantic{b}\sigma = \false  
		\end{cases}\\
		\ssemantics{\ifR{b}{c}{c'}}&\defeq\semantic{c}(\sigma,\mu | \semantic{b}\sigma\neq \false )\padd{\mu(\semantic{b}\sigma\neq \false )}\\
		&\quad\semantic{c'}(\sigma,\mu | \semantic{b}\sigma=\false)\\
		\ssemantics{\whileD{b}{c}}&\defeq\ssemantics{\ifDs{b}{(c;\ \whileD{b}{c})}} \\
		\ssemantics{\whileR{b}{c}}&\defeq\ssemantics{\ifRs{b}{(c;\ \whileR{b}{c})}}
	\end{align*}
	\caption{Programming Language Semantics\label{fig:full-semantics}}
	\label{semantic}
\end{figure}
Unlike PSL, which defines its loop
semantics somewhat informally, ours
enables direct
mechanisation (in Isabelle/HOL).

\section{Logic}\label{sec:logic}

\subsection{Assertions}\label{sec:assertions}

The assertions of our logic include those of PSL, which we
extend with the certainty assertion $\Local{e_r}$ while extending
the uniform distribution assertion $\Uniform{e_d}{e_r}$ by
allowing the set to be
specified by an expression~$e_d$ (rather than
a constant as in PSL). The free variables of an expression~$e$ are denoted $\FV{e}$.
The domain of distribution $\mu$ over memories, written
$\dom{\mu}$, is the set of random variables in
the memories in the support of $\mu$.
$\AP$ denotes the set of atomic assertions.

For a random variable expression~$e_r$, $\Local{e_r}$ asserts
that $e_r$ evaluates to true in every memory consistent with the
current configuration, i.e.\ it holds with absolute certainty.
Note that the set of random variable expressions~$e_r$ can
accommodate all standard assertions from classical Hoare logic.

\begin{definition}[Atomic assertion semantics]\label{defn:atomic-assertions}
	\begin{align*}
		\semantic{\Local{e_r}} &= \{(\sigma,\mu)\ |\ \forall m \in \supp{\mu}.\ \ \semantic{e_r}(\sigma,m) = \true\}\\
		\semantic{\Uniform{e_d}{e_r}} &= \{(\sigma,\mu)\ |\ \FV{e_r}\union\FV{e_d}\subseteq \dom{\sigma}\union\dom{\mu}\\
		&\qquad\qquad\qquad\text{and }\Unif{\vset{\semantic{e_d}\sigma}} = \semantic{e_r}(\sigma,\mu)\}
	\end{align*}
\end{definition}

The assertion $\Uniform{e_d}{e_r}$ asserts that the evaluation
of random variable expression $e_r$ yields the uniform
distribution over the set denoted by the deterministic expression $e_d$ when evaluated in the current deterministic memory, where the $\vset{}$
function is used to retrieve that denotation after evaluating
$e_d$ (\cref{vset}).
We require the expression~$e_d$ to be deterministic as otherwise this assertion can introduce contradictions (e.g.\ if the
set expression instead denoted a truly random set including possible sets $\{1,2\}$ and $\{0\}$, then $e_r$ will not be uniformly distributed on any set).  

From PSL our logic inherits its other assertions with Kripke resource monoid. The assertions $\always$ (which holds always),
$\never$ (which never holds), and connectives
$\land$, $\lor$, $\imply$ have their standard meaning. The separation logic~\cite{EarlySL}
connectives $*$ is separating
conjunction, here used to assert probabilistic independence;
and $\sepi$ is separating implication. See \ifExtended
\cref{app:defns}
\else
Appendix A.2 of the extended version of the paper on Arxiv
\fi
for details.

Note that $\Local{P}\land \Local{Q}$ is equivalent to $\Local{P\land Q}$, but $\Local{a=1}$ $\lor \Local{a=2}$ is different to $\Local{a=1\lor a=2}$:  the former asserts that either $a$ is always 1 or $a$ is always 2 (stronger); the latter asserts that always $a$ is either 1 or 2 (weaker).

We also write $\DD{x}$ to abbreviate $\Local{x = x}$, which asserts that the variable $x$ is in the domain of the partial configuration. Any distribution of $x$ satisfies this assertion.

Finally, we introduce several useful propositions about assertions implication. They are very useful in the verification and reflect our interplay between classical and probabilistic independence reasoning, especially the last one.

\begin{proposition}\label{prop}
	\begin{align}
		&\semA (\phi\sep\psi)\land\eta \imply (\phi \land \eta)\sep \psi \text{ ,where } \semA \phi \imply \DD{\FV{\eta}\inter \RV}\\
		&\semA (\phi\sep\psi) \imply (\phi\land\psi)\\
		&\semA \Uniform{S}{e} \land\Local{f \text{is a bijection from } S \text{to } S'} \imply \Uniform{S'}{f(e)} \\
		&\semA (\Local{\phi \land \psi}) \imply (\Local{\phi}\land\Local{\psi})\\
		&\semA (\Local{\phi}\land\Local{\psi}) \imply (\Local{\phi \land \psi})\\
		&\semA \Uniform{S}{e}\imply \Local{e\in S}\\
		&\semA \Uniform{S}{e} \land \Local{e=e'}\imply \Uniform{S}{e'}\\
		&\semA \Local{x = e\land x\notin \FV{e'}} \land \DD{e} \sep \DD{e'} \implies \DD{x}\sep \DD{e'}
	\end{align}
	\vspace*{-0.5cm}
	\[
	\semA \begin{array}{l} \begin{array}{@{}l@{}} \Local{\forall a,b\in S, c,d\in S'.f(a,c) = f(b,d) \imply a=b \land c=d}\land
			\Uniform{S}{x}\sep \Uniform{S'}{e'} \end{array} \\
		\imply \Uniform{S\times_f S'}{f(x,e')}, \text{ where } S\times_f S' = \{f(a, b)~|~a\in S \land b \in S'\}  \end{array} (9)
	\]
	
	\label{implication}
\end{proposition}

The first two are inherited from PSL. 
The third one generalises a similar proposition of PSL~\cite{PSL} over possibly different sets $S$ and $S'$.
The fourth and fifth
show the equivalence of $\land$ whether
inside or outside the certain assertions.
The sixth shows the straightforward consequence that if $e$ is uniformly distributed over set $S$, then the value of $e$ must be in $S$. The seventh shows two expressions satisfy the same distribution if they are certainly equal. The eighth shows if we know that $e$ is independent of $e'$ and we know another variable $x = e$ additionally, we can conclude that $x$ is also independent of $e'$ if $x$ is not a free variable in $e'$. 

The last one also generalises a proposition of PSL~\cite{PSL} by leveraging $\Local{\cdot}$ conditions: it restricts binary function $f$ by requiring it to produce different outputs when
given two different pairs of inputs. In practice, we will use this lemma letting $f$ be the concatenation function on two arrays where $S$ is a set of possible arrays with the same length. We conclude the concatenated array satisfies the uniform distribution on $S$ times $S'$ if those premises hold.

\subsection{Judgements and Rules}
\label{sec:rule}

The judgements~$\triple{\phi}{c}{\psi}$ of our program logic are simple Hoare logic
correctness statements, in which~$c$ is a program command and
$\phi$ and $\psi$ are preconditions and postconditions respectively.

\begin{definition}[Judgement Validity]
	Given two assertions $\phi, \psi$ and a program command $c$, a judgement $\{\phi\}c\{\psi\}$ is valid if for all configuration $(\sigma,\mu)$ satisfying $(\sigma,\mu)\semA \phi$, we have $\semantic{c}(\sigma,\mu)\semA\psi$, denoted $\triple{\phi}{c}{\psi}$.
	\label{triple}
\end{definition}

\begin{figure}
	\begin{mathpar}
		\noindent\infer[\textsc{RAssign}]{\phi \in \AP}{ \triple{\phi[e_r/x_r]}{\assign{x_r}{e_r}}{\phi}}
		
		\infer[\textsc{RSample}]
		{\triple{\Local{\EI(f,S,S')}}{\rassign{x_r}{S}}{\Uniform{S'}{f(x_r)}}}{}
		
		\infer%
		{ \triple{\Local{\phi \land b \neq \false}}{c}{\Local{\psi}} \qquad
			\triple{\Local{\phi \land b = \false} }{c'}{\Local{\psi}}
		}
		{\triple{\Local{\phi}}{\ifR{b}{c}{c'}}{\Local{\psi}}}\textsc{RCond}
		
		\infer{ \triple{\phi}{\ifRs{b}{c}}{\phi}}{\triple{\phi}{\whileR{b}{c}}{\phi \land \Local{b = \false}}}\textsc{RLoop}
		
		\infer[\textsc{Unif-Idp}]
		{ \FV{a} \inter \MV{c} = \emptyset \qquad b\notin\FV{a} \qquad
			\triple{\Local{a\in A}\sep Q \land\Local{P}}{c}{\Uniform{S}{b}}
		}
		{\triple{\Local{a\in A}\sep Q \land\Local{P}}{c}{(\DD{a}*\Uniform{S}{b})} }
	\end{mathpar}
	\caption{Rules capturing the interplay of classical and probabilistic reasoning.
		\label{rules}}
\end{figure}

Our logic inherits all of PSL's original rules~\cite{PSL} (See \ifExtended
\cref{Orules}
\else
Appendix A.3 of the extended version of the paper on Arxiv
\fi 
for details); many of them use the $\Local{\cdot}$
assertion to encode equality tests, which were encoded instead
in PSL primitively.

\cref{rules} depicts the rules of our logic that embody its new reasoning
principles,
and support the requirements listed at the end
of \cref{sec:motivation}.
The random assignment rule~\textsc{RAssign} has the classical Hoare logic form. It requires
the postcondition~$\phi$ is atomic to avoid unsound derivations, such as $\NOtriple{0=0*0=0}{x=0}{x=x*x=x}$.

As mentioned in \cref{sec:nondet}, the \textsc{RSample} rule is another embodiment
of the general principle underlying the design of our logic, of classical and
probabilistic reasoning enhancing each other. Specifically, it allows us to deduce
when a randomly sampled quantity~$f(x_r)$ (a function~$f$ applied to a random
variable~$x_r$) is uniformly distributed over set~$S'$
when the random variable~$x_r$ was uniformly sampled over set~$S$. It is especially
useful when~$S$ is itself random. It relies on the function~$f$
\emph{evenly partitioning} the input set~$S$ into~$S'$, as defined below.

\begin{definition}[Even Partition]\label{defn:even-partition}
  Given two sets $S, S'$ and a function $f$, we
  say that $f$ \emph{evenly partitions} $S$ into $S'$ if and only if $S' = \{f(s)|s\in S\}$ and there exists an integer $k$ such that
  $\forall s'\in S'. \left|\{s\in S | f(s) = s'\}\right| = k$.
  In this case we write $\EI(f,S,S')$.
	\label{even}
\end{definition}

\textsc{RSample} allows reasoning over random choices beyond
original PSL~\cite{PSL}, and in particular dynamic random
sampling from truly random sets. For example, at line $\linenum{10}$ of
\cref{fig:motivation}, we have $\Local{\EI(f,S,S')}$ where
$f = ~\%~8,~S = \{0\cdots m\},~S' = \{0\cdots 7\}$. 
Letting $k = m/8$ with the above definition, we can prove the pre-condition implies $\Local{\EI(f,S,S')}$.
Note that if $m=9$ then $\Local{\EI(f,S,S')}$ will not hold because we cannot find $k$.
The existence of $k$ makes sure that $S$ can be \emph{evenly} partitioned to $S'$ by $f$. Also, from our new random sample rule~\textsc{RSample}, one can obtain PSL's original rule by letting $S'=S$ and $f=(\lambda x.\ x)$.

Besides PSL's random conditional rule,
we also include the \textsc{RCond} rule for random
conditions that operate over certainty assertions~$\Local{\cdot}$.
It is in many cases more applicable
because it does not require the branching
condition to be independent of the precondition and, while
it reasons only over certainty assertions, other conditions
can be added by applying the \textsc{Const} rule (\cref{Orules}).
The new random loop rule~\textsc{RLoop} is straightforward,
requiring proof of the invariant~$\phi$ over a random
conditional.

The final new rule \textsc{Unif-Idp} unifies two methods to prove the independence of an
algorithm's output~$b$ from its input~$a$:
it says that if
given any arbitrary distribution of $a$ we can always prove that the result~$b$ is uniformly distributed, then $a$ and $b$ are independent because the distribution of $a$ does not influence $b$, where $\MV{c}$ is the variables $c$ may write to (same as PSL's definition). It is useful for programs that consume their secrets by random choice at runtime (e.g. \cref{fig:motivation} we verified in \cref{sec:nondet} and the Oblivious Sampling algorithm~\cite{OlyaSampling} we verify in \ifExtended
\cref{sec:OlyaSampling}).
\else
Appendix C.2 of the extended version of the paper on Arxiv).
\fi

As an example, we used this rule between line $\linenum{10}$ and line $\linenum{11}$ in \cref{fig:motivation} by letting $a = (O,~A[1-S[i]])$ and $P, Q$ be the other information in the assertion before line $\linenum{10}$. The first premise of the rule is true because these two lines of code never modify $O$ and $A[1-S[i]]$. The second premise is also trivially true. The third premise is proved by the \textsc{RSample} and \textsc{RAssign} rules. This yields the conclusion that $O$ and $A[1-S[i]]$ are independent of $A[S[i]]$.

Note that the pre-condition $\Local{a\in A}\sep Q \land\Local{P}$ appears in both premise and conclusion of the rule. Considering the \textsc{Weak} rule (\cref{Orules}; aka the classical consequence rule), when the precondition is in the premise, we want it be as strong as it can so that the premise is easier to be proved. When it is in conclusion, we want it be as weak as it can so that the conclusion is more useful. These two requirements guide us to design the rule with two free assertions connected by $\land$ and $\sep$ respectively so that it is very flexible. If we change the pre-condition to $\DD{a}$ (deleting $A,P,Q$), this rule is still sound (which can be proved by letting $A$ be the universe set and $P,Q$ be true) but much less applicable.

\subsection{Soundness}
\begin{theorem}
	All the rules in \cref{rules} and \cref{Orules}, plus the other original PSL rules~\cite{PSL} omitted from~\cref{Orules}, are sound, i.e. are valid according to \cref{triple}.
\end{theorem}

We formalised our logic and proved it sound in Isabelle/HOL (see Supplemental Material). It
constitute 7K lines of Isabelle and required approx.\
8 person-months to complete. Some of our Isabelle proofs follow PSL's pen-and-paper proofs
but we also found several problems in PSL's definitions and
proofs. We briefly discuss those now, to highlight the
value and importance of machine-checked proofs for establishing
the soundness of program logics.

\subsection{Oversights in original PSL}\label{unsound}

Our machine-checked proofs identified various oversights 
in the pen-and-paper formalisation of original PSL~\cite{PSL}.
We fixed them either by modifying specific
definitions or by finding an alternative---often much more
complicated, but sound---proof strategy.

PSL~\cite{PSL} defines the notion of when
a formula~$\phi$ is \emph{supported} ($\SP$), requiring that
for any deterministic memory $\sigma$, there exists a
distribution over random
variable memories $\mu$ such that if $(\sigma, \mu')\semA \phi$, then $\mu \sqsubseteq \mu'$ (meaning that $\mu$ is a marginal distribution of $\mu'$ where $\dom{\mu} \subseteq \dom{\mu'}$)~\cite[Definition 6]{PSL}.

This definition aims to restrict the assertions used in
PSL's original rule for random conditionals~\cite[rule \textsc{RCond} of Figure 3]{PSL}, but it is not strong enough. All the assertions satisfy it because $\mu$ can always be instantiated with the unit distribution over the empty
memory $\unit{\emptyset \rightarrow \Val}$, $\sqsubseteq$ all others. This means the second example in their
paper~\cite[Example 2]{PSL} is a counterexample to their
rule for random conditionals because there is not any non-supported assertion.

We fixed this
by altering their definition of $\SP$. 
Note that simply excluding the empty memory case is not enough to fix this problem.
Instead, we have \cref{def:sp} and our Isabelle proofs ensure its soundness. It does not have a big impact on adjusting the proofs strategy of relevant rules.

\begin{definition}[Supported] An assertion $\phi$ is Supported ($\SP$) if for any deterministic memory $\sigma$, there exists a
	randomised memory $\mu$ such that if $(\sigma, \mu')\semA \phi$, then $\mu \sqsubseteq \mu'$ and $(\sigma, \mu)\semA \phi$.
	\label{def:sp}
\end{definition}

Additionally,
key lemmas that underpin PSL's soundness argument turned
out to be true, but not for the reasons stated in their proofs~\cite[Lemmas 1 and 2, Appendix B]{PSL}.
PSL's Lemma 1 proof has mistakes in the implication case. The second sentence of the implication case said, ``there exists a distribution $\mu''$ such that…". However $\mu''$ may not exists because $\mu$ and $\mu'$ may disagree on some variables in $\FV{\phi1,\phi2}$.
PSL's Lemma 2 proof also has mistakes. They said ``we have $(\sigma1, \mu1) \semA \eta$" on the third line of proof but this is not true because $\sigma1$ may not equal $\sigma$ (the domain of $\sigma1$ could be smaller than $\sigma$). The actual proof of these needs a different strategy which formalized in Isabelle by us.

Without mechanising the soundness of our program logic,
it is unlikely we would have uncovered these issues. This
shows the vital importance of mechanised soundness proofs.

\section{Case Studies}\label{sec:examples}

We applied our program logic
to verify the obliviousness of four 
non-trivial oblivious algorithms: the
Melbourne Shuffle~\cite{Melbourne},
Oblivious Sampling~\cite{OlyaSampling},
Path ORAM~\cite{pathORAM} and 
Path Oblivious Heap~\cite{pathOheap}. The details are in \ifExtended
\cref{app:case}.
\else
Appendix C of the extended version of the paper on Arxiv.
\fi

While these proofs are manual, each took less than a person-day to complete,
except for Path Oblivious Heap,
which took approx.\ 2 days of proof effort.

To our knowledge, 
the Melbourne Shuffle,
Oblivious Sampling,
and Path Oblivious Heap
have never been formally verified as each
requires the combination of features that our approach
uniquely supports.
Path ORAM
has received some formal verification~\cite{sahai2020verification,leungtowards} (see later
in \cref{sec:related-work})
and also comes with
an informal but rigorous proof of security~\cite{pathORAM}. We verified it
to show that our logic can indeed encode existing
rigorous security arguments.

In practice we need to distinguish the public memory locations and private locations where we assume any access to public memory locations is visible to attackers. So we will add ghost codes to record all those access in an array ``Trace" and finally we aim to prove the array is independent of secrets.

The Melbourne Shuffle \cite{Melbourne} (see \ifExtended
\cref{sec:shuffle})
\else
Appendix C.1 of the extended version of the paper on Arxiv)\fi
is an effective oblivious shuffling algorithm used in cloud storage
and also a basic building block for other higher-level algorithms (e.g. oblivious sampling \cite{OlyaSampling}).
Its operation is
non-trivial, including re-arranging array elements with dummy values and other complexities.
Its verification makes heavy use of classical reasoning because, while it is probabilistic,
its memory access pattern is deterministic in the absence of failure. 

Oblivious sampling \cite{OlyaSampling} (see \ifExtended
\cref{sec:OlyaSampling})
\else
Appendix C.2 of the extended version of the paper on Arxiv)\fi is another important building block having
applications in differential privacy, oblivious data analysis and machine learning.
The algorithm obliviously samples from a data set, by producing a uniformly-distributed memory access pattern, and
includes random and secret-dependent looping and if-statements, plus dynamic random choices
(shuffling on a truly probabilistic array).
Thus the interplay between classical and probabilistic reasoning that our logic provides
is essential to verifying its security.

Path ORAM \cite{pathORAM} (see \ifExtended
\cref{sec:pathORAM})
\else
Appendix C.3 of the extended version of the paper on Arxiv)\fi is a seminal oblivious RAM algorithm
with practical
efficiency, providing general-purpose oblivious storage. Path oblivious heap
(\ifExtended
\cref{sec:pathOheap})
\else
Appendix C.4 of the extended version of the paper on Arxiv)\fi is inspired by Path ORAM and the two share the same idea:
using a binary tree with a random and virtual location table to store secret data,
where the mappings between each physical and virtual location are always independent of each other and of the memory access pattern. Thus probabilistic independence is crucial to express and
prove these algorithms' key invariants.

\section{Related Work}\label{sec:related-work}

Our program logic naturally extends PSL~\cite{PSL} non-trivially, including support
for classical reasoning, dynamic
random choice, improved support for  random
conditionals, random loops, and random assignments. Our mechanisation of PSL
identified and fixed a number of soundness issues (see \cref{unsound}).

Its unique synergy of classical and probabilistic
independence reasoning means  our program logic is more
expressive not only than PSL but also prior probabilistic Hoare
logics, such as~\cite{pHL_Den}, VPHL~\cite{VPHL} and Easycrypt's
pRHL~\cite{Easycrypt}. 

Probabilistic coupling (supported by pRHL and Easycrypt~\cite{Easycrypt}) is another popular way for proving the security of probabilistic algorithms. It does so by proving the output distribution is equal between any pair of different secret inputs, witnessed by a bijection probabilistic coupling for each probabilistic choice. However, for dynamic random choice, the bijection probabilistic coupling may not exist or may even be undefined (e.g. \cref{fig:motivation} and \cite{OlyaSampling}). Sometimes, finding the correct coupling can be far more challenging than
proving the conclusion directly via probabilistic independence. Indeed, the original
informal security
proofs of our case studies \cite{OlyaSampling, pathORAM, pathOheap, Melbourne} all use probabilistic independence to argue their obliviousness, instead of coupling.

Other program logics or type systems for verifying obliviousness also exist.
For example, ObliCheck \cite{Oblicheck} and $\lambda_{\text{OADT}}$ \cite{OADT} can be used to check or prove  obliviousness but only for deterministic algorithms.
$\lambda_{\mathit{obliv}}$~\cite{RObliv} is a type system
for a functional language for
proving obliviousness
of probabilistic algorithms but it
forbids branching on secrets, which is prevalent in many
oblivious algorithms including those we consider in \cref{sec:examples}.
It also forbids outputting a probabilistic value (and all other values influenced by it) more than once.
Our approach
suffers no such restriction.

Path ORAM has received some verification attention~\cite{sahai2020verification,leungtowards}. \cite{sahai2020verification} reason about this algorithm but in a non-probabilistic model, instead representing it as a nondeterministic transition system, and apply model counting to prove a security property about it. Their property says that for any observable output, there is a sufficient number of inputs to hide which particular input would have produced that output. This specification seems about the best that can be achieved for a nondeterministic model of the algorithm, but would also hold for an implementation that used biased choices (which would necessarily reveal too much of the input). Ours instead says that for each input the output is identically distributed, and would not be satisfied for such a hypothetical implementation. Nonetheless, it would be interesting to compare the strengths and weaknesses of their
complementary approach to ours.  \cite{leungtowards} recently proposed to verify this algorithm in Coq, but as far as we are aware ours is the first verification of Path ORAM via a probabilistic program logic.

Other recent work extends PSL in different ways. One \cite{computationalPSL} extended PSL to computational security, but it cannot deal with loops (neither deterministic nor probabilistic) so their target algorithms are very different to ours. Lilac\cite{Lilac} also uses separating conjunction to model probabilistic independence. Crucially, it supports reasoning about conditional probability and conditional independence; \cite{NominalPSL} validated the design decisions of Lilac. However, Lilac's programming language is functional whereas ours is imperative; Lilac does not support random loops or dynamic random choice, which are essential for our aim.

IVL \cite{IVL} reasons about probabilistic programs with nondeterminism. In doing so it supports classical reasoning (e.g. for the nondeterministic parts) and probabilistic reasoning for the probabilistic parts. Our logic reasons only about probabilistic programs (with no nondeterminism) but allows using classical reasoning to reason about parts of the probabilistic program, and for the classical and probabilistic reasoning styles to interact and enhance each other.


\section{Conclusion and Future Work}

We presented the first program logic that, to our knowledge, is
able to verify the obliviousness of real-world foundational probabilistic oblivious
algorithms 
whose implementations combine challenging features
like dynamic random choice and secret- and random-variable-dependent
control flow.
Our logic harnesses the interplay between
classical and probabilistic reasoning, is situated atop PSL~\cite{PSL}, and proved sound in Isabelle/HOL.
We applied it to several challenging case studies,
beyond the reach of prior approaches. 

\bibliographystyle{splncs04}
\bibliography{references}

\begin{thebibliography}{10}
\providecommand{\url}[1]{\texttt{#1}}
\providecommand{\urlprefix}{URL }
\providecommand{\doi}[1]{https://doi.org/#1}

\bibitem{ConstantTime}
Almeida, J.B., Barbosa, M., Barthe, G., Dupressoir, F., Emmi, M.: Verifying
  constant-time implementations. In: USENIX Security Symposium. vol.~16, pp.
  53--70 (2016)

\bibitem{QIFBook}
Alvim, M.S., Chatzikokolakis, K., McIver, A., Morgan, C., Palamidessi, C.,
  Smith, G.: The Science of Quantitative Information Flow. Springer (2020)

\bibitem{Easycrypt}
Barthe, G., Dupressoir, F., Gr{\'e}goire, B., Kunz, C., Schmidt, B., Strub,
  P.Y.: Easycrypt: A tutorial. Foundations of Security Analysis and Design VII:
  FOSAD 2012/2013 Tutorial Lectures pp. 146--166 (2014)

\bibitem{PSL}
Barthe, G., Hsu, J., Liao, K.: A probabilistic separation logic. Proc. ACM
  Program. Lang.  \textbf{4}(POPL) (dec 2019). \doi{10.1145/3371123},
  \url{https://doi.org/10.1145/3371123}

\bibitem{Stash}
Bittau, A., Erlingsson, U., Maniatis, P., Mironov, I., Raghunathan, A., Lie,
  D., Rudominer, M., Kode, U., Tinnes, J., Seefeld, B.: Prochlo: Strong privacy
  for analytics in the crowd. In: Proceedings of the 26th Symposium on
  Operating Systems Principles. p. 441–459. SOSP '17, Association for
  Computing Machinery, New York, NY, USA (2017). \doi{10.1145/3132747.3132769},
  \url{https://doi.org/10.1145/3132747.3132769}

\bibitem{FaCT}
Cauligi, S., Soeller, G., Johannesmeyer, B., Brown, F., Wahby, R.S., Renner,
  J., Gr{\'e}goire, B., Barthe, G., Jhala, R., Stefan, D.: Fact: a dsl for
  timing-sensitive computation. In: Proceedings of the 40th ACM SIGPLAN
  Conference on Programming Language Design and Implementation. pp. 174--189
  (2019)

\bibitem{RObliv}
Darais, D., Sweet, I., Liu, C., Hicks, M.: A language for probabilistically
  oblivious computation. Proc. ACM Program. Lang.  \textbf{4}(POPL) (dec 2019).
  \doi{10.1145/3371118}, \url{https://doi.org/10.1145/3371118}

\bibitem{rawORAM}
Fletcher, C.W., Ren, L., Kwon, A., van Dijk, M., Stefanov, E., Devadas, S.:
  {RAW} path {ORAM:} {A} low-latency, low-area hardware {ORAM} controller with
  integrity verification. {IACR} Cryptol. ePrint Arch. p.~431 (2014),
  \url{http://eprint.iacr.org/2014/431}

\bibitem{Kripke}
Galmiche, D., M\'{e}ry, D., Pym, D.: The semantics of {BI} and resource
  tableaux. Mathematical Structures in Computer Science  \textbf{15}(6),
  1033–1088 (2005). \doi{10.1017/S0960129505004858}

\bibitem{SquareRoot}
Goldreich, O., Ostrovsky, R.: Software protection and simulation on oblivious
  rams. J. ACM  \textbf{43}(3),  431–473 (may 1996).
  \doi{10.1145/233551.233553}, \url{https://doi.org/10.1145/233551.233553}

\bibitem{10.1007/978-3-642-22012-8_46}
Goodrich, M.T., Mitzenmacher, M.: Privacy-preserving access of outsourced data
  via oblivious ram simulation. In: Aceto, L., Henzinger, M., Sgall, J. (eds.)
  Automata, Languages and Programming. pp. 576--587. Springer Berlin
  Heidelberg, Berlin, Heidelberg (2011)

\bibitem{191010}
Gruss, D., Spreitzer, R., Mangard, S.: Cache template attacks: Automating
  attacks on inclusive {Last-Level} caches. In: 24th USENIX Security Symposium
  (USENIX Security 15). pp. 897--912. USENIX Association, Washington, D.C. (Aug
  2015),
  \url{https://www.usenix.org/conference/usenixsecurity15/technical-sessions/presentation/gruss}

\bibitem{pHL_Den}
den Hartog, J.I.: Verifying probabilistic programs using a hoare like logic.
  In: Thiagarajan, P.S., Yap, R. (eds.) Advances in Computing Science ---
  ASIAN'99. pp. 113--125. Springer Berlin Heidelberg, Berlin, Heidelberg (1999)

\bibitem{Hoare}
Hoare, C.A.R.: An axiomatic basis for computer programming. Commun. ACM
  \textbf{12}(10),  576–580 (oct 1969). \doi{10.1145/363235.363259},
  \url{https://doi.org/10.1145/363235.363259}

\bibitem{hsu2017}
Hsu, J.: Probabilistic couplings for probabilistic reasoning. Ph.D. thesis,
  University of Pennsylvania (2017)

\bibitem{Modern}
Katz, J., Lindell, Y.: Introduction to Modern Cryptography Second Edition.
  Chapman, Hall/CRC, 2nd edn. (2014)

\bibitem{10.5555/2095116.2095129}
Kushilevitz, E., Lu, S., Ostrovsky, R.: On the (in)security of hash-based
  oblivious ram and a new balancing scheme. In: Proceedings of the Twenty-Third
  Annual ACM-SIAM Symposium on Discrete Algorithms. pp. 143--156. SODA '12,
  Society for Industrial and Applied Mathematics, USA (2012)

\bibitem{computationalPSL}
Lago, U.D., Davoli, D., Kapron, B.M.: On separation logic, computational
  independence, and pseudorandomness (extended version) (2024),
  \url{https://arxiv.org/abs/2405.11987}

\bibitem{SGX}
Lee, S., Shih, M.W., Gera, P., Kim, T., Kim, H., Peinado, M.: Inferring
  fine-grained control flow inside {SGX} enclaves with branch shadowing. In:
  26th USENIX Security Symposium (USENIX Security 17). pp. 557--574. USENIX
  Association, Vancouver, BC (Aug 2017),
  \url{https://www.usenix.org/conference/usenixsecurity17/technical-sessions/presentation/lee-sangho}

\bibitem{leungtowards}
Leung, H., Ringer, T., Fletcher, C.W.: Towards formally verified path oram in
  coq  (2023), available online:
  \url{https://dependenttyp.es/pdf/oramproposal.pdf}

\bibitem{Lilac}
Li, J.M., Ahmed, A., Holtzen, S.: Lilac: A modal separation logic for
  conditional probability. Proc. ACM Program. Lang.  \textbf{7}(PLDI) (jun
  2023). \doi{10.1145/3591226}, \url{https://doi.org/10.1145/3591226}

\bibitem{NominalPSL}
Li, J.M., Aytac, J., Johnson-Freyd, P., Ahmed, A., Holtzen, S.: A nominal
  approach to probabilistic separation logic. In: Proceedings of the 39th
  Annual ACM/IEEE Symposium on Logic in Computer Science. LICS '24, Association
  for Computing Machinery, New York, NY, USA (2024).
  \doi{10.1145/3661814.3662135}, \url{https://doi.org/10.1145/3661814.3662135}

\bibitem{ObliVM}
Liu, C., Wang, X.S., Nayak, K., Huang, Y., Shi, E.: Oblivm: A programming
  framework for secure computation. In: 2015 IEEE Symposium on Security and
  Privacy. pp. 359--376 (2015). \doi{10.1109/SP.2015.29}

\bibitem{Last-Level}
Liu, F., Yarom, Y., Ge, Q., Heiser, G., Lee, R.B.: Last-level cache
  side-channel attacks are practical. In: 2015 IEEE Symposium on Security and
  Privacy. pp. 605--622 (2015). \doi{10.1109/SP.2015.43}

\bibitem{PHANTOM}
Maas, M., Love, E., Stefanov, E., Tiwari, M., Shi, E., Asanovic, K.,
  Kubiatowicz, J., Song, D.: Phantom: practical oblivious computation in a
  secure processor. In: Proceedings of the 2013 ACM SIGSAC Conference on
  Computer \& Communications Security. p. 311–324. CCS '13, Association for
  Computing Machinery, New York, NY, USA (2013). \doi{10.1145/2508859.2516692},
  \url{https://doi.org/10.1145/2508859.2516692}

\bibitem{PC-Security}
Molnar, D., Piotrowski, M., Schultz, D., Wagner, D.: The program counter
  security model: Automatic detection and removal of control-flow side channel
  attacks. In: Information Security and Cryptology-ICISC 2005: 8th
  International Conference, Seoul, Korea, December 1-2, 2005, Revised Selected
  Papers 8. pp. 156--168. Springer (2006)

\bibitem{EarlySL}
O'Hearn, P., Reynolds, J., Yang, H.: Local reasoning about programs that alter
  data structures. In: Fribourg, L. (ed.) Computer Science Logic. pp. 1--19.
  Springer Berlin Heidelberg, Berlin, Heidelberg (2001)

\bibitem{Leakage}
Ohrimenko, O., Costa, M., Fournet, C., Gkantsidis, C., Kohlweiss, M., Sharma,
  D.: Observing and preventing leakage in mapreduce. In: Proceedings of the
  22nd ACM SIGSAC Conference on Computer and Communications Security. p.
  1570–1581. CCS '15, Association for Computing Machinery, New York, NY, USA
  (2015). \doi{10.1145/2810103.2813695},
  \url{https://doi.org/10.1145/2810103.2813695}

\bibitem{Melbourne}
Ohrimenko, O., Goodrich, M.T., Tamassia, R., Upfal, E.: The melbourne shuffle:
  Improving oblivious storage in the cloud. In: Esparza, J., Fraigniaud, P.,
  Husfeldt, T., Koutsoupias, E. (eds.) Automata, Languages, and Programming.
  pp. 556--567. Springer Berlin Heidelberg, Berlin, Heidelberg (2014)

\bibitem{VPHL}
Rand, R., Zdancewic, S.: Vphl: A verified partial-correctness logic for
  probabilistic programs. Electronic Notes in Theoretical Computer Science
  \textbf{319},  351--367 (12 2015). \doi{10.1016/j.entcs.2015.12.021}

\bibitem{sahai2020verification}
Sahai, S., Subramanyan, P., Sinha, R.: Verification of quantitative
  hyperproperties using trace enumeration relations. In: Computer Aided
  Verification: 32nd International Conference, CAV 2020, Los Angeles, CA, USA,
  July 21--24, 2020, Proceedings, Part I 32. pp. 201--224. Springer (2020)

\bibitem{OlyaSampling}
Sasy, S., Ohrimenko, O.: Oblivious sampling algorithms for private data
  analysis. In: Proceedings of the 33rd International Conference on Neural
  Information Processing Systems. Curran Associates Inc., Red Hook, NY, USA
  (2019)

\bibitem{IVL}
Schr\"{o}er, P., Batz, K., Kaminski, B.L., Katoen, J.P., Matheja, C.: A
  deductive verification infrastructure for probabilistic programs. Proc. ACM
  Program. Lang.  \textbf{7}(OOPSLA2) (oct 2023). \doi{10.1145/3622870},
  \url{https://doi.org/10.1145/3622870}

\bibitem{pathOheap}
Shi, E.: Path oblivious heap: Optimal and practical oblivious priority queue.
  Cryptology ePrint Archive, Paper 2019/274 (2019),
  \url{https://eprint.iacr.org/2019/274},
  \url{https://eprint.iacr.org/2019/274}

\bibitem{Oblicheck}
Son, J., Prechter, G., Poddar, R., Popa, R.A., Sen, K.: {ObliCheck}: Efficient
  verification of oblivious algorithms with unobservable state. In: 30th USENIX
  Security Symposium (USENIX Security 21). pp. 2219--2236. USENIX Association
  (Aug 2021),
  \url{https://www.usenix.org/conference/usenixsecurity21/presentation/son}

\bibitem{pathORAM}
Stefanov, E., Dijk, M.V., Shi, E., Chan, T.H.H., Fletcher, C., Ren, L., Yu, X.,
  Devadas, S.: Path oram: An extremely simple oblivious ram protocol. J. ACM
  \textbf{65}(4) (apr 2018). \doi{10.1145/3177872},
  \url{https://doi.org/10.1145/3177872}

\bibitem{OADT}
Ye, Q., Delaware, B.: Oblivious algebraic data types. Proc. ACM Program. Lang.
  \textbf{6}(POPL) (jan 2022). \doi{10.1145/3498713},
  \url{https://doi.org/10.1145/3498713}

\bibitem{Opaque}
Zheng, W., Dave, A., Beekman, J.G., Popa, R.A., Gonzalez, J.E., Stoica, I.:
  Opaque: An oblivious and encrypted distributed analytics platform. In: 14th
  USENIX Symposium on Networked Systems Design and Implementation (NSDI 17).
  pp. 283--298. USENIX Association, Boston, MA (Mar 2017),
  \url{https://www.usenix.org/conference/nsdi17/technical-sessions/presentation/zheng}

\end{thebibliography}

\appendix

\section{Ancillary Definitions}\label{app:defns}
Here we spell out in detail how our programming language and assertion semantics,
for those inherited from PSL, is defined. These
definitions follow those from PSL~\cite{PSL}. The difference has been introduced in \cref{sec:pre} and \cref{sec:logic}.

\subsection{Security Definition and Verification by Approximation}\label{sec:security}

As we said, many oblivious algorithms have a very small failure probability. Now we state 
the formal security property that we target,
known in this paper as $\epsilon$\emph{-Statistical Secrecy}, and the process connecting it with our logic.
Our security definition is familiar from standard cryptographic security definitions~\cite{Modern}. It is
a straightforward relaxation of the following
observation. Suppose we have an algorithm that operates over
a secret but whose output reveals nothing about that secret. Without loss of
generality assume the secret is a single bit: either 0 or 1.
Then, assuming the secret is chosen uniformly (i.e.\ with equal probability)
over $\{0,1\}$, an attacker who is only able to observe the output of this
algorithm can guess the value of the secret correctly
with probability no more than $\frac{1}{2}$.

Statistical secrecy relaxes this to allow the attacker a small margin of
advantage, $\epsilon$, permitting them to guess
correctly with probability at most~$\frac{1}{2} + \epsilon$.

\begin{definition}[$\epsilon$-Statistical Secrecy] \label{def:security}
	Suppose a probabilistic algorithm $f$ accepts some secret data $S$ as input and produces some information $f(S)$ which can be observed by some attacker. We say $f$ satisfies $\epsilon$\emph{-statistical secrecy} if and only if
	for any two different secrets $S_1$ and $S_2$, if we choose $S$ from the uniform distribution on $\{S_1, S_2\}$ and then reveal $f(S)$ to the attacker, then the attacker's probability of correctly
	guessing whether $S$ was $S_1$ or $S_2$ is at most $\frac{1}{2} + \epsilon$.
	\label{almost}
\end{definition}

We note that this property is an instance of a quantitative
information flow (QIF)~\cite{QIFBook} guarantee about the algorithm
against an attacker whose prior over the secret is uniform and who
is modelled by the gain function in which a correct guess of the
secret is assigned the value 1 and an incorrect guess the value 0,
guaranteeing that the change in the attacker's gain is at
most $\epsilon$.

The security of oblivious algorithms can be expressed as a simple instance
of $\epsilon$-statistical security relative to an attacker
that can directly observe the memory access pattern, as follows.

\begin{definition}[Obliviousness]\label{defn:obliviousness}
	Suppose an algorithm $f$ takes some secret data $S$ and produces some memory access pattern $f(S)$. We say $f$ is oblivious iff for any two different secrets $S_1$ and $S_2$ with the same length $n$, if we choose $S$ from the uniform distribution on $\{S_1, S_2\}$ and then reveal $f(S)$, then the attacker has no greater than probability $1/2 + g(n)$ to guess the value of $S$ correctly, where $g(n)$ is a negligible function of $n$.
\end{definition}

Our approach to proving statistical secrecy is inspired by
informal proofs of obliviousness for existing algorithms~\cite{pathOheap, OlyaSampling, pathORAM, Melbourne},
in which reasoning proceeds by ``factoring out'' the sources of
imperfection in the algorithm to consider an implicitly perfect,
hypothetical version of the algorithm. Reasoning proceeds by
arguing rigorously but informally
that the hypothetical version is perfectly oblivious and, therefore,
that the original algorithm is oblivious. This last step is
performed by quantifying the difference between the original
imperfect
algorithm and the hypothetical perfect version and using this
distance to bound the degree of imperfection and to argue that it
is indeed negligible. The measure of difference used is
\emph{Statistical Distance} (sometimes called \emph{Total Variation Distance}~\cite{hsu2017}).

\begin{definition}[Statistical Distance]
	Given two distributions $p$ and $q$ over the same sample space $S$, we write the statistical distance (also called total variation distance) between $p$ and $q$ as $\SD(p, q) = \frac{1}{2}\sum_{s\in S} |p(s) - q(s)|$
\end{definition}

\begin{lemma}\label{lem:sd}
	Suppose there is a distribution $D$ and an algorithm $f$ such that for any input $S$, the statistical distance between $f(S)$ and $D$ is smaller or equal to $\epsilon$, then $f$ satisfies $\epsilon$-statistical secrecy.
\end{lemma}

This lemma is a well-known fact and its proof is relegated to \cref{app:proofs}.

Thus our approach to verifying an imperfect oblivious algorithm is to build a
\emph{perfectly oblivious approximation}, such that for all inputs
the statistical distance between the two is bounded by a
negligible amount, then we use our logic to verify that the \emph{perfectly oblivious approximation} leak no information.

\subsection{Assertions}

\begin{definition}[Assertions]\label{defn:assertions}
	Assertions, $\phi$, $\psi$ etc. are defined as:\\
	$\phi, \psi ::= p\ |\ \always\ |\ \never\ |\ \phi\land\psi\ |\ \phi\lor\psi\ |\ \phi\imply\psi\ |\ \phi*\psi\ |\ \phi\sepi\psi$\\
	where $\AP\ni p::= \Local{e_r}\ |\ \Uniform{e_d}{e_r}$
\end{definition}

Assertion's semantics is given via a
partial Kripke resource monoid~\cite{Kripke}.

\begin{definition}[Kripke resource monoid~\cite{Kripke}]
	\ \\
	A (partial) Kripke resource monoid consists of a set $M$, a partial binary operation $\circ: M \times M \rightharpoonup M$, an element $e \in M$, and a pre-order $\sqsubseteq$ on $M$ such that:\\
	\hspace*{0.4cm}$e$ is the identity, namely $\forall x \in M.\ x = e\circ x = x\circ e$\\
	\hspace*{0.4cm}$\circ$ is associative: namely $x\circ(y\circ z) = (x\circ y)\circ z$ (both sides can be undefined)\\
	\hspace*{0.4cm}$\circ$ is compatible with the pre-order: namely $x\circ x' \sqsubseteq y\circ y'$ if both sides are defined, $x \sqsubseteq y$, and $x' \sqsubseteq y'$
\end{definition}

As in PSL~\cite{PSL}, $M$ is instantiated to be the set of
all \emph{partial} memories, $e$ is the empty memory,
$\circ$ combines two disjoint memories, and
$m \sqsubseteq m'$ means $m$ is a sub-memory of $m'$. We write $(m\circ m')\downarrow$ to say $(m\circ m')$ is defined.

For any $S\subseteq \RV$, the set of all corresponding memories is denoted $\RanMx{S}\defeq S\rightarrow \Val$. Moreover, for any $\mu\in\DD{\RanMx{S}}$, the \emph{domain} of $\mu$, written $\dom{\mu}$, is $S$.

The empty memory (whether deterministic or random) is defined as $\emptyset \rightarrow \Val$.

Given two disjoint sets $S, S'\subseteq \RV$ and two distributions of memories over them, $\mu_S\in\DD{\RanMx{S}}$ and $\mu_{S'}\in\DD{\RanMx{S'}}$, we define the product of these two distributions: $\mu_S \Dproduct \mu_{S'} = \lambda m.\ \mu_S(p_S(m))\cdot \mu_{S'}(p_{S'}(m))$, where $m$ is a memory (of type $\RV\rightarrow\Val$), and $p_S(m)$ returns the sub-map of $m$ over $S$. For any $v\notin S$, $p_S(m)(v)$ is undefined.

Given $S'\subseteq S$ and a distribution $\mu$ over $S$, we define $\pi_{S, S'}$ to yield the distribution over $S'$ (the smaller domain), $\pi_{S, S'}(\mu)\defeq \\ \lambda m'.\ \Sig{m\in\{m|p_{S'}(m) = m'\}}{\mu(m)}$, where $p_{S'}(m)$ takes the sub-map of $m$ over $S'$. For example, let $S = \{x, y, z\}$, and $ S' = \{x\}$, let $\mu$ be a distribution over $S$ talking about the distribution of the values of those 3 variables, then $\pi_{S, S'}(\mu)$ is the corresponding distribution of the value of variable $x$. We will omit $S$ later because it is always $\dom{\mu}$.

When we evaluate expressions on partial memories, the results may be undefined if some variables in the expression are not in the domain of the partial memories.

\begin{definition}[Instantiating Kripke resource monoid]
	\ \\
	$M$ is the set of partial configurations $\DetM[S]\times\DD{\RanM[T]}$, where $S$ and $T$ are subsets of $\DV$ and $\RV$ respectively.\\
	$e\defeq(\emptyset \rightarrow \Val,~ \unit{\emptyset \rightarrow \Val})$\\
	$(\sigma,\mu)\circ (\sigma', \mu')\defeq
	\begin{cases}
		(\sigma\cup\sigma',\mu\Dproduct \mu') & :\sigma = \sigma' \text{ on }\dom{\sigma}\cap\dom{\sigma'}\\
		&\ \ \text{ and }\dom{\mu}\cap\dom{\mu'} = \emptyset\\    
		\text{undefined} & :\text{otherwise}
	\end{cases}$\\
	$(\sigma,\mu)\sqsubseteq (\sigma', \mu') \text{ iff }
	\begin{cases}
		& \dom{\sigma}\subseteq\dom{\sigma'}\text{ and }\sigma = \sigma' \text{ on }\dom{\sigma}\\    
		& \dom{\mu}\subseteq\dom{\mu'}\text{ and }\mu = \pi_{\dom{\mu}}(\mu')
	\end{cases}$\\
	Given these definitions, $(M,\circ,e,\sqsubseteq)$ is a Kripke resource monoid.
\end{definition}

Let $m\models \phi$ denote when partial configuration $m$ satisfies assertion $\phi$. Then the assertion semantics is defined as follows~\cite{PSL}.

\begin{definition}[Assertion Semantics]
	\ \\
	For any set $\AP$ of atomic assertions, an interpretation function $\semantic{-}:\AP \rightarrow 2^M$, and a Kripke resource monoid $(M,\circ,e,\sqsubseteq)$ such that if $m\in \semantic{p}$ and $m\sqsubseteq m'$, then $m'\in \semantic{p}$, the semantics of non-atomic proposition assertions is defined as follows:
	\begin{align*}
		&m\models p&&\text{iff } m\in \semantic{p}\\
		&m\models \always&& \text{always}\\
		&m\models\ \never&& \text{never}\\
		&m\models \phi \land \psi&&\text{iff } m\models \phi \text{ and } m\models \psi\\
		&m\models \phi \lor \psi&&\text{iff } m\models \phi \text{ or } m\models \psi\\
		&m\models \phi \rightarrow \psi&&\text{iff for all } m\sqsubseteq m', m'\models \phi \text{ implies } m'\models \psi\\
		&m\models \phi * \psi&&\text{iff exist } m_1, m_2.~~ (m_1\circ m_2)\downarrow\text{ and } m_1\circ m_2 \sqsubseteq m \\ 
		&\ &&\text{ and } m_1\models \phi\text{ and } m_2\models \psi\\
		&m\models \phi \sepi \psi&&\text{iff for all } m'\models \phi,~~ (m\circ m')\downarrow \text{ implies } m\circ m'\models \psi
	\end{align*}
	\label{AS}
\end{definition}

All assertions satisfy the Kripke monotonicity property: if $m\semA\phi$ and $m \sqsubseteq m'$, then $m'\semA\phi$.
We write $\semA \phi$ when $\phi$ holds in all partial configurations.

\subsection{Inherited Rules and Auxiliary Functions}

\cref{auxfun} defines the straightforward auxiliary functions for computing sets of
variables read/modified by a command. \cref{Orules} shows rules inherited from PSL\cite{PSL}; many of them use the $\Local{\cdot}$
assertion to encode equality tests, which were encoded instead
in PSL primitively.

\begin{figure*}
	\begin{mathpar}
		\ReadV{\assign{x_r}{e}} \defeq \FV{e}\qquad \ReadV{\rassign{x_r}{e}}\defeq\FV{e} \qquad \ReadV{\ifD{b}{c}{c'}} \defeq \ReadV{c}\union\ReadV{c'}\qquad \ReadV{\whileD{b}{c}} \defeq \ReadV{c}\\
		\ReadV{c;c'} \defeq \ReadV{c}\union(\ReadV{c'}-\WV{c}) \qquad\ReadV{\ifR{b}{c}{c'}} \defeq \ReadV{c}\union\ReadV{c'} \union \FV{b}\qquad \ReadV{\whileR{b}{c}} \defeq \ReadV{c}\union\FV{b}\\
		
		\noindent\makebox[\linewidth]{\rule{17.5cm}{0.4pt}}
		
		\WV{c;c'} \defeq \WV{c}\union(\WV{c'}-\ReadV{c}) \qquad \WV{\ifD{b}{c}{c'}} \defeq \WV{c}\inter\WV{c'}\\
		\WV{\assign{x_r}{e}} \defeq \{x_r\} - \FV{e}\qquad \WV{\rassign{x_r}{e}}\defeq \{x_r\} - \FV{e} \qquad\WV{\ifR{b}{c}{c'}} \defeq (\WV{c}\inter\WV{c'}) - \FV{b}\\
		
		\noindent\makebox[\linewidth]{\rule{17.5cm}{0.4pt}}
		
		\MV{\assign{x_d}{e}} \defeq \{x_d\} \qquad \MV{c;c'} \defeq \MV{c}\union\MV{c'} \qquad \MV{\whileD{b}{c}} \defeq \MV{c} \qquad \MV{\ifD{b}{c}{c'}} \defeq \MV{c}\union\MV{c'}\\
		\MV{\assign{x_r}{e}} \defeq \{x_r\}\qquad \MV{\rassign{x_r}{e}}\defeq \{x_r\} \qquad \MV{\whileR{b}{c}} \defeq \MV{c}\qquad\MV{\ifR{b}{c}{c'}} \defeq \MV{c}\union\MV{c'}\\
		
		\noindent\makebox[\linewidth]{\rule{17.5cm}{0.4pt}}
	\end{mathpar}
	\caption{Auxiliary Functions for Rules}
	\label{auxfun}
\end{figure*}

\begin{figure*}
	\begin{mathpar}
		\noindent\infer[\textsc{DAssign}]{}{ \triple{\phi[e_d/x_d]}{\assign{x_d}{e_d}}{\phi}}
		
		\infer[\textsc{Skip}]
		{\triple{\phi}{\sskip}{\phi}}{}
		
		\infer[\textsc{Seqn}]
		{ \triple{\phi}{c}{\psi} \qquad
			\triple{\psi}{c'}{\eta}
		}
		{\triple{\phi}{c;c'}{\eta}}
		
		\infer[\textsc{Weak}]
		{ \triple{\phi}{c}{\psi} \qquad
			\semA\phi' \imply \phi \qquad \semA\psi \imply \psi'
		}
		{\triple{\phi'}{c}{\psi'}}
		
		\infer[\textsc{Const}]
		{ \triple{\phi}{c}{\psi} \qquad
			\FV{\eta}\inter\MV{c} = \emptyset
		}
		{\triple{\phi \land \eta}{c}{\psi \land \eta}}
		
		\infer[\textsc{DCond}]
		{ \triple{\phi \land \Local{b \neq \false}}{c}{\psi} \qquad
			\triple{\phi \land \Local{b = \false} }{c'}{\psi}
		}
		{\triple{\phi}{\ifD{b}{c}{c'}}{\psi}}
		
		\infer[\textsc{True}]
		{\triple{\always}{c}{\always}}{}
		
		\infer[\textsc{DLoop}]{ \triple{\phi\land \Local{b \neq \false}}{c}{\phi}}{\triple{\phi}{\whileD{b}{c}}{\phi \land \Local{b = \false}}}
		
		\infer[\textsc{RDCond}]
		{ \triple{\phi \land \Local{b \neq \false}}{c}{\psi} \qquad
			\triple{\phi \land \Local{b = \false} }{c'}{\psi} \qquad
			\semA\phi \imply (\Local{b \neq \false} \lor \Local{b = \false})
		}
		{\triple{\phi}{\ifR{b}{c}{c'}}{\psi}}
		
		\infer[\textsc{RCond}]
		{ \triple{\phi \sep \Local{b \neq \false}}{c}{\psi\sep \Local{b \neq \false}} \qquad
			\triple{\phi \sep \Local{b = \false} }{c'}{\psi\sep \Local{b = \false}} \qquad
			\psi \in \SP
		}
		{\triple{\phi\sep \DD{b}}{\ifR{b}{c}{c'}}{\psi\sep \DD{b}}}
		
		\infer[\textsc{Conj}]
		{ \triple{\phi_1}{c}{\psi_1} \qquad
			\triple{\phi_2}{c}{\psi_2}
		}
		{\triple{\phi_1 \land \phi_2}{c}{\psi_1 \land \psi_2}}
		
		\infer[\textsc{Case}]
		{ \triple{\phi_1}{c}{\psi_1} \qquad
			\triple{\phi_2}{c}{\psi_2}
		}
		{\triple{\phi_1 \lor \phi_2}{c}{\psi_1 \lor \psi_2}}
		
		\infer[\textsc{RCase}]
		{ \triple{\phi \sep \Local{b \neq \false}}{c}{\psi\sep \Local{b \neq \false}} \qquad
			\triple{\phi \sep \Local{b = \false} }{c}{\psi\sep \Local{b = \false}} \qquad
			\psi \in \SP
		}
		{\triple{\phi\sep \DD{b}}{c}{\psi\sep \DD{b}}}
		
		\infer[\textsc{Frame}]
		{ \triple{\phi}{c}{\psi} \qquad
			\FV{\eta}\inter\MV{c} = \emptyset \qquad
			\FV{\psi}\subseteq T \union \ReadV{c} \union \WV{c} \qquad
			\semA\phi \imply \DD{T \union \ReadV{c}} 
		}
		{\triple{\phi \sep \eta}{c}{\psi \sep \eta}}
		
	\end{mathpar}
	\caption{Rules inherited from PSL~\cite{PSL}.}
	\label{Orules}
\end{figure*}

\section{Omitted Proofs}\label{app:proofs}
This proof of \cref{lem:sd} appears in \cref{proof:sd}.

\begin{figure*}
	\begin{proof}
		We note firstly that
		for any two inputs $S_1, S_2$, the statistical distance between $f(S_1)$ and $f(S_2)$ is at most $2\epsilon$, since
		the distance between each $f(S_i)$ and~$D$ is at most $\epsilon$ and by the transitivity of statistical distance. \\
		\ \\
		Let function $g : E \rightarrow [0,1]$ model the attacker's strategy of guessing the result, where $E$ is the set of all possible observations and $g(e)$ represents the probability that the attacker guesses $S_1$ under observation $e$; otherwise the attacker guesses $S_2$ with the probability $1-g(e)$. Then the overall probability of a correct guess is:
		$\frac{1}{2}\sum_{e\in E}(f(S_1)(e) \cdot g(e)) + \frac{1}{2}\sum_{e\in E}(f(S_2)(e) \cdot (1-g(e)))$\\
		\ \\
		Because both $f(S_1)(e)$ and $f(S_2)(e)$ are smaller or equal to $\max(f(S_1)(e),f(S_2)(e))$, the above expression is smaller or equal to $\frac{1}{2}\sum_{e\in E}\max(f(S_1)(e),f(S_2)(e))$.\\
		\ \\
		Then because\\
		$\sum_{e\in E}\max(f(S_1)(e),f(S_2)(e)) + \sum_{e\in E}\min(f(S_1)(e),f(S_2)(e)) = \sum_{e\in E}f(S_1)(e)+f(S_2)(e) = 2$ and\\
		$\sum_{e\in E}\max(f(S_1)(e),f(S_2)(e)) - \sum_{e\in E}\min(f(S_1)(e),f(S_2)(e)) = \sum_{e\in E}|f(S_1)(e)-f(S_2)(e)| = 2\SD(a,b)$,\\
		we have $\sum_{e\in E}\max(f(S_1)(e),f(S_2)(e)) = (2 + \SD(f(S_1),f(S_2)))/2\le 1+2\epsilon$.\\
		\ \\
		Thus we have the correct probability is smaller or equal to $\frac{1}{2}\sum_{e\in E}\max(f(S_1)(e),f(S_2)(e)) \le \frac{1}{2}+\epsilon$
	\end{proof}
	\caption{Proof of \cref{lem:sd}}
	\label{proof:sd}
\end{figure*}

\begin{figure}
		\[
		\begin{array}{@{}r@{\ \ \ \ \ \ }l}
			&\hspace*{-0.4cm}\mathsf{synthetic}(S,O,n):\\
			\linenum{1}&\rassign{A[0]}{\{0,1,2,\cdots,7\}};\\
			\linenum{2}&\rassign{A[1]}{\{0,1,2,\cdots,7\}};~\dassign{i}{0};\\
			\linenum{3}&\whileN{n>i}{}\\
			\linenum{4}&\hspace*{0.4cm} \dassign{O}{O + A[S[i]]};\\
			\linenum{5}&\hspace*{0.4cm} \dassign{m}{8};~\dassign{j}{0};\\
			\linenum{6}&\hspace*{0.4cm} \whileN{A[S[i]]>j}{}\\
			\linenum{7}&\hspace*{0.8cm} \dassign{m}{2\times m};~\dassign{j}{j+1};\ \\
			\linenum{8}&\hspace*{0.8cm} \ifNs{(j+S[i])~\%~3 == 0}{}\ \\
			\linenum{9}&\hspace*{1.2cm} \dassign{j}{j+1};\ \\
			\linenum{10}&\hspace*{0.4cm} \rassign{t}{\{1,2,3,\cdots,m\}};\\
			\linenum{11}&\hspace*{0.4cm} \dassign{A[S[i]]}{t~\%~8};\\
			\linenum{12}&\hspace*{0.4cm} \dassign{i}{i+1};
		\end{array}
		\]
		\caption{\label{fig:motivation0}A synthetic motivating example algorithm.}
\end{figure}

\section{Case Studies}\label{app:case}

\subsection{The Melbourne Shuffle}\label{sec:shuffle}
The Melbourne Shuffle \cite{Melbourne} is a probabilistic oblivious algorithm that takes an array (database) $I$ and a target permutation $\pi$ as its secret input. Its job is to
shuffle $I$ according to the permutation $\pi$, placing the
shuffled data into the output array $O$. To do so, it makes
use of a (larger) temporary array $T$. The algorithm is
designed to be oblivious despite the access patterns of $I, T$ and $O$ being observable to attackers.

In practice, this algorithm is designed to facilitate e.g.\
oblivious training of machine learning models in the cloud:
the arrays $I$ and $O$ are held locally by a client; shuffling
is performed by a server whose memory is~$T$. All communication
between the client and server is encrypted; however, there might
be spies who can nonetheless observe the access patterns to
the three arrays (e.g.\ via cache side channels). Moreover,
the cloud provider might be malicious and so might directly
observe the contents of array~$T$ to learn the secret data or
the desired permutation. For this reason, the algorithm
keeps the contents of~$T$ encrypted.

Our verification abstracts away from the client-server
communication and encryption, effectively assuming the latter
is perfect. We capture the threat model above by adding
\emph{ghost code} (i.e.\ code that records information
during an algorithm without affecting the execution of the
algorithm) to record the attacker's observations of the
memory access patterns to arrays~$I$, $T$ and~$O$.
Throughout this paper, ghost code is written in blue.

The algorithm appears in \cref{fig:shuffle-alg}.
Its ghost code records the memory access pattern as a
sequence~$Trace$ of accesses. Each access is a tuple
$(\mathit{oper},A,k)$ where $\mathit{oper}$ is the operation
performed by the access (e.g. read/write/copy), $A$ is the array
being accessed (e.g.~$I$), and $k$ is the position of the array
being accessed by the operation. For operations that copy
the entire contents of one array to another, $k$ will denote
the source array.
Given a sequence $A$, we define $\size{A}$ as the length of $A$, $A + a$ as the result of appending $a$ to $A$ and $A \concat B$ to concatenate two arrays~$A$ and~$B$.

As when the algorithm was proposed~\cite{Melbourne},
target permutations~$\pi$ are represented as functions from a piece of data to its expected index~\cite{Melbourne}.
For example, if the array is $[x, y, z]$ and $\pi(x) = 1, \pi(y) = 0, \pi(z) = 2$, then after a shuffle pass, the output should be $[y, x, z]$. Without loss of generality, the input
array~$I$ is assumed to contain distinct elements.

We write $F(I)$ to represent the set of all possible $\pi$ given a array $I$.

\begin{figure}
	\begin{algorithmic}
		\Function{shuffle}{$I,\pi,O$}
		\State $\ghost{\dassign{Trace}{[]}}$
		\State $\rassign{\pi_1}{F(I)}$
		\State $\dassign{T}{[]}$
		\State $\call{shuffle\_pass}{I,T,\pi_1,O,\textcolor{blue}{Trace}}$
		\State $\dassign{I}{O};\record{("copy", "O", "I")}$
		\State $\call{shuffle\_pass}{I,T,\pi,O,\textcolor{blue}{Trace}}$
		\EndFunction
	\end{algorithmic}
	\caption{The Melbourne Shuffle\label{fig:shuffle-alg}}
\end{figure}

The Melbourne shuffle hides the target permutation~$\pi$
by performing the shuffle in two passes. In the first pass,
it randomly selects a permutation~$\pi_1$ from the
set of all possible permutations of~$I$ and shuffles~$I$
into~$T$ according to~$\pi_1$. This ensures that the
items in~$T$ are at unpredictable locations. It is therefore
now safe to shuffle~$T$ into~$O$ according to the desired
permutation~$\pi$. 

This algorithm has a small chance of failing. In particular,
because the arrays~$I$ and~$O$ are necessarily smaller
than~$T$, the internal \textsf{shuffle\_pass} function can
fail for certain choices of its given permutation.
Fortunately the probability of failure is a negligible
function of the size of~$I$~\cite{Melbourne}. 

We construct a perfectly oblivious approximation of \cref{fig:shuffle-alg} by modifying the random selection of $\pi_1$
so that rather than choosing from all permutations~$F(I)$
it instead selects from the set of permutations under which
the algorithm doesn't fail. The resulting algorithm and its
proof is shown in \cref{fig:mel}.
Letting $\pi_I$ denote the unique
permutation that describes the contents of the input array~$I$,
this set of permutations under which the algorithm doesn't
fail is $S_p(\pi_I) \union S_p(\pi)$, where $S_p$ is
defined as follows.

Let $V(\pi,i,j) = \{v~|~i\le\pi(v)<j\} $ denote $i$th to $j$th values in the permutation $\pi$. Then $S_p$ is a function
that takes a permutation whose length is $n$, and returns the set of permutations for which the input array uniquely described by $\pi$
can be shuffled without failure. $S_p$ is defined as the
set of all $\pi$' such that:\\
$\size{\pi} = \size{\pi'} = n$ and $\forall i\in\{0,1,...,\sqrt{n}-1\}$, \\
\hspace*{1.5cm}$\size{V(\pi, \sqrt{n}\cdot i, \sqrt{n} \cdot (i+1)) \cap V(\pi', \sqrt{n} \cdot i, \sqrt{n}*(i+1))} \le p\cdot\log(n)$.

This definition of $S_p$ is symmetric about $\pi$ and $\pi'$,
so we have that $\forall \pi, \pi'.~~ \pi\in S_p(\pi')\iff \pi'\in S_p(\pi)$. Thus, the Melbourne shuffle succeeds if and only if $\pi_1 \in S_p(\pi_I) \union S_p(\pi)$~\cite[Lemma 4.4]{Melbourne}. 

The statistical distance between \cref{fig:mel} and \cref{fig:shuffle-alg} is at most a negligible function of~$n$~\cite[Lemma 4.3]{Melbourne}. We can thus verify \cref{fig:shuffle-alg}
by proving
perfect obliviousness of the approximation in \cref{fig:mel}.

We do so by applying our logic to prove that it produces a
fixed, deterministic memory access pattern. Letting
$\call{TS}{``I",``T",``O",n}$ denote the unique memory
access pattern produced by calling
$\call{shuffle\_pass}{I,T,\pi,O}$ for any $\pi$, we prove
the memory access pattern of \cref{fig:mel} is
$\call{TS}{``I",``T",``O",n}+(``copy", ``O", ``I")\concat\call{TS}{``I",``T",``O",n}$, as shown in the final postcondition.

\begin{figure*}
	\begin{algorithmic}
		\Function{shuffle'}{$I,\pi,O$}\\
		$\assert{\{\Local{\size{I} = n = \size{\Set{I}} \land \forall v\in I.~~ I[\pi_I(v)] = v}\}}$
		\State $\ghost{\dassign{Trace}{[]}}$\\
		$\assert{\{\Local{Trace = [] \land \size{I} = n = \size{\Set{I}}\land \forall v\in I.~~ I[\pi_I(v)] = v}\}}$
		\State $\rassign{\pi_1}{S_p(\pi)\cap S_p(\pi_I)}$\\
		$\assert{\{\Local{Trace = []\land \size{I} = n = \size{\Set{I}}\land (\forall v\in I.~~ I[\pi_I(v)] = v)}\sep\Uniform{\pi_1}{S_p(\pi)\cap S_p(\pi_I)}\}}$\\
		$\assert{\{\Local{Trace = []\land \size{I} = n = \size{\Set{I}}\land (\forall v\in I.~~ I[\pi_I(v)] = v) \land \pi_1 \in S_p(\pi)\cap S_p(\pi_I)}\}}$
		\State $\dassign{T}{[]}$\\
		$\assert{\{\Local{Trace = []\land \size{I} = n = \size{\Set{I}}\land (\forall v\in I.~~ I[\pi_I(v)] = v) \land T=[] \land \pi_1 \in S_p(\pi)\cap S_p(\pi_I)}\}}$
		\State $\call{shuffle\_pass}{I,T,\pi_1,O,\textcolor{blue}{Trace}}$\\
		$\assert{\{\LocalS(Trace = \call{TS}{"I","T","O",n}\land \size{O} = n = \size{\Set{O}} \land(\forall v\in O.~~ O[\pi_1(v)] = v) \ \land}$\\
		$\;\;\;\assert{\pi_1 \in S_p(\pi)\cap S_p(\pi_I))\}}$
		\State $\dassign{I}{O};\record{("copy", "O", "I")}$\\
		$\assert{\{\LocalS(Trace = \call{TS}{"I","T","O",n}+("copy", "O", "I")\land \size{I} = n = \size{\Set{I}}\ \land}$\\
		$\;\;\;\assert{(\forall v\in I.~~ I[\pi_1(v)] = v) \land \pi \in S_p(\pi_1))\}}$
		\State $\call{shuffle\_pass}{I,T,\pi,O,\textcolor{blue}{Trace}}$\\
		$\assert{\{\LocalS(Trace = \call{TS}{"I","T","O",n}+("copy", "O", "I")\concat\call{TS}{"I","T","O",n} \ \land}$\\
		$\;\;\;\assert{\size{O} = n = \size{\Set{O}} \land(\forall v.~~ O[\pi(v)] = v))\}}$
		\EndFunction
	\end{algorithmic}
	\caption{Verification of the perfect Melbourne shuffle. $\Set{A}$ is the set of values in array~$A$.
		\label{fig:mel}}
\end{figure*}

The precondition of \cref{fig:mel} states that the initial
array~$I$ contains~$n$ unique elements, and that $\pi_I$
does indeed describe its contents correctly.

This proof relies on the inner function \textsf{shuffle\_pass}
adhering to the following specification.

\[
\begin{array}{c}
	\assert{\{\Local{Trace = X\land \size{I} = n = \size{\Set{I}} \land (\forall v\in I.~~ I[\pi_1(v)] = v) \land \pi \in S_p(\pi_1)}\}}\\
	\hspace*{2.6cm}\call{shuffle\_pass}{I,T,\pi,O,\textcolor{blue}{Trace}}\\
	\assert{\{\Local{Trace = X\concat\call{TS}{"I","T","O",n} \land \size{O} = n  = \size{\Set{O}} \land (\forall v\in O.~~ O[\pi(v)] = v)}\}}
\end{array}
\]

Note that this is a classical Hoare logic specification, and
that \textsf{shuffle\_pass} is a deterministic algorithm.
Therefore this specification
can be proved in classical Hoare logic~\cite{Hoare} and so
is omitted.  Importantly, the postcondition of this specification
states also that the shuffle is correctly performed:
$\forall v\in O.~~ O[\pi(v)] = v$. This is crucial as otherwise
the two calls to \textsf{shuffle\_pass} cannot be chained
together (as the precondition for the second call relies on
this assumption). Thus the obliviousness of this algorithm
depends on its correctness. This explains the vital importance
of being able to mix classical correctness reasoning with
probabilistic reasoning, as supported by our logic via the
$\Local{\cdot}$ assertions and associated rules (\cref{rules}).

\subsection{Oblivious Sampling}\label{sec:OlyaSampling}

Our second case study is oblivious random sampling~\cite{OlyaSampling}. This algorithm makes use of an oblivious shuffling
primitive, which can be implemented using the Melbourne shuffle
from \cref{sec:shuffle}. As alluded to in
\cref{sec:shuffle}, oblivious sampling is an important part of
oblivious training algorithms for machine learning, e.g.\ to
randomly sample mini-batches.

The algorithm~\cite{OlyaSampling} takes a secret database (an array) $D$ of size~$n$ as input, and will output several arrays ($s[0], s[1], \ldots , s[k]$) where each contains $m$ pieces of independently sampled data from $D$. Moreover, $n = m \cdot k$. The memory access pattern of the database $D$, temporary array $S$ and the returned arrays $s[\ldots]$ are observable to attackers. As in \cref{sec:shuffle}, we abstract from encryption and client-server communication.

The perfectly oblivious approximation of the algorithm, with
ghost code to capture the observable memory access pattern,
is shown in \cref{rewriteS}. The source of imperfection in
the original algorithm is the oblivious shuffle primitive,
which here has been replaced by uniform random choice
$\rassign{D}{\Perm{D}}$ over the set~$\Perm{D}$ of permutations
of~$D$ (and likewise for~$S$), thereby removing the source of imperfection. We
prove the resulting approximation is perfectly oblivious.

Following the original algorithm~\cite{OlyaSampling}, the
array $SWO$ is a two-dimensional array of booleans of size $k \times n$,
which is randomly chosen from the set $X(n,m)$ which
contains all such arrays such that  $\forall i\in\{1\cdots k\}$, the number of $\true$ in $SWO[i][1\cdots n]$ is $m$. 
Unlike in the Melbourne Shuffle (\cref{sec:shuffle}, where arrays are indexed from 1), the arrays in \cref{rewriteS}
are indexed from 0~\cite{OlyaSampling,Melbourne}.

\begin{figure}
	\flushleft
	$\ghost{\dassign{Trace}{[]}};$\\
	$\rassign{D}{\Perm{D}};~\record{\text{oblishuffle}(n)}$\\
	$\rassign{SWO}{X(n,m)}~;$
	$\dassign{S}{[]}~$;
	$\dassign{j}{1}~$;
	$\dassign{l}{1}~$;\\
	$\dassign{e}{D[1]};~\record{(``read", ``D", 1)}$\\
	$\dassign{e_{next}}{D[1]};~\record{(``read",``D", 1)}$\\
	$\whileR{l<n+1}{}$\\
	\hspace*{0.4cm} $\dassign{i}{1}$\\
	\hspace*{0.4cm} $\whileR{i<k+1}{}$\\
	\hspace*{0.8cm} $\ifRs{SWO[i][j]}{}$\\
	\hspace*{1.2cm} $\dassign{S}{S+(e,i)};$\\
	\hspace*{1.2cm} $\record{(``write", ``S", \size{S}+1)};$\\
	\hspace*{1.2cm} $\dassign{l}{l+1};$\\
	\hspace*{1.2cm} $\dassign{e_{next}}{D[l]};$\\
	\hspace*{1.2cm} $\record{("read","S",l)}$\\
	\hspace*{0.8cm} $\dassign{i}{i+1};\ $\\
	\hspace*{0.4cm} $\dassign{e}{e_{next}};\ \dassign{j}{j+1};$\\
	$\rassign{S}{\Perm{S}};\ \record{\text{oblishuffle}(\size{S})}$\\
	$\dassign{s}{[[],[],\cdots,[]]}; // \text{k empty arrays}$\\
	$\dassign{p}{1};$\\
	$\whileR{p<\size{S}+1}{}$\\
	\hspace*{0.4cm} $\dassign{(e, i)}{S[p]};$\\
	\hspace*{0.4cm} $\record{(``read", ``S", p)}$\\
	\hspace*{0.4cm} $\dassign{s[i]}{s[i]+e};$\\
	\hspace*{0.4cm} $\record{(``write", ``s", i, \size{s[i]}+1)}$\\
	\hspace*{0.4cm} $\dassign{p}{p+1};$
	
	\caption{Rewritten Sampling Algorithm\label{rewriteS}}
\end{figure}

The verification of the sampling algorithm has two main parts.
The first part is from the beginning of the algorithm to the end of the first loop. In this part, the memory access trace is deterministic, and so we prove that as a certainty
via $\Local{\cdot}$ reasoning. At this point, we have
that the trace is a deterministic value (captured by the
predicate $\inv{Trace}$), plus some certain
information about $\second{S}$, which is essential for the
verification that follows.

The second part of the verification covers the remaining code,
which shuffles $S$ and then produces a memory access pattern
which is a deterministic function of (the shuffled) $\second{S}$.
We thus prove perfect obliviousness by proving
that the overall memory access trace
is uniformly distributed (and thus independent of secrets).

Then we introduce the detailed reasoning of the proof, sketched in \cref{proofS}. The reasoning is as follows.

\begin{figure*}
	\begin{minipage}{1.1\textwidth}
		\begin{small}
			\flushleft
			\assert{$\{\Local{n = m \times k}\}$}\\
			$\ghost{\dassign{Trace}{[]}};$\\
			$\rassign{D}{\Perm{D}};~\record{\text{oblishuffle}(n)}$\\
			$\rassign{SWO}{X(n,m)}~;$
			$\dassign{S}{[]}~$;
			$\dassign{j}{1}~$;
			$\dassign{l}{1}~$;\\
			$\dassign{e}{D[1]};~\record{(``read", ``D", 1)}$\\
			$\dassign{e_{next}}{D[1]};~\record{(``read",``D", 1)}$\\
			\assert{$\{\LocalS(Trace = [ \text{oblishuffle}(n), (``read, ``D", 1), (``read",``D", 1)] \land n = m \times k \land S = [] \land j = l = 1 \land e = e_{next} = D[1] \land$}\\
			\hspace{0.6cm} \assert{$D\in \Perm{D}\land SWO \in X(n,m))\}$}\\
			\centerline{Start Unif-Idp rule on \assert{$SWO$} and \assert{$Trace$}}
			$\whileR{l<n+1}{}$\\
			\hspace{0.4cm} \assert{Loop Invariant:}\\
			\hspace*{0.4cm} \assert{$\{ \LocalS(n = m \times k \land D\in \Perm{D}\land SWO \in X(n,m) \land l = \size{S} + 1 = (\size{Trace}-1)/2 = \Count{\true, SWO[1..k][1..j-1]}+1 \land \inv{Trace}\land(\forall x.\ 0<x\le k \implies \Count{\true, SWO[x][1..j-1]}=\Count{x,\second{S}})\land \size{Trace}\%2 = 1\land l\le n+1)\}$}\\
			\hspace*{0.4cm} $\dassign{i}{1}$\\
			\hspace*{0.4cm} $\whileR{i<k+1}{}$\\
			\hspace{0.8cm} \assert{Loop Invariant:}\\
			\hspace{0.8cm} \assert{$\{ \LocalS(n = m \times k \land D\in \Perm{D}\land SWO \in X(n,m) \land i \le k + 1\land l = \size{S} + 1 = (\size{Trace}-1)/2 = \Count{\true, SWO[1..k][1..j-1]}+\Count{\true,SWO[1..i-1][j]}+1 \land\inv{Trace}\land (\forall x.\ 0<x< i \implies \Count{\true, SWO[x][1..j]}=\Count{x,\second{S}})\land (\forall x.\ i\le x\le k \implies \Count{\true, SWO[x][1..j-1]}=\Count{x,\second{S}}))\}$\\}
			\hspace*{0.8cm} $\ifRs{SWO[i][j]}{}$\\
			\hspace*{1.2cm} $\dassign{S}{S+(e,i)};$\\
			\hspace*{1.2cm} $\record{(``write", ``S", \size{S}+1)};$\\
			\hspace*{1.2cm} $\dassign{l}{l+1};$\\
			\hspace*{1.2cm} $\dassign{e_{next}}{D[l]};$\\
			\hspace*{1.2cm} $\record{("read","S",l)}$\\
			\hspace*{0.8cm} $\dassign{i}{i+1};\ $\\
			\hspace*{0.4cm} $\dassign{e}{e_{next}};\ \dassign{j}{j+1};$\\
			\assert{$\{\LocalS(\size{Trace} = 2n+3\land \inv{Trace} \land \size{S} = n\land(\forall x.\ 0<x\le k \implies \Count{x,\second{S}}=m))\}$\\}
			$\rassign{S}{\Perm{S}};$\\
			\assert{$\{\LocalS(\size{Trace} = 2n+3\land \inv{Trace} \land \size{S} = n\land(\forall x.\ 0<x\le k \implies \Count{x,\second{S}}=m)) \land \Uniform{\Perm{Se}}{\second{S}}\}$\\}
			$\record{\text{oblishuffle}(\size{S})}$\\
			$\dassign{s}{[[],[],\cdots,[]]}; // \text{k of empty arrays}$\\
			$\dassign{p}{1};$\\
			\assert{$\{\LocalS(\invv{Trace,S}\land \size{S} = n \land p = \size{Trace}/2-n-1\land (\forall x.\ 0<x\le k \implies \Count{x,\second{S}}=m))\land \Uniform{\Perm{Se}}{\second{S}}\}$}\\
			\centerline{Start Const rule with \assert{$\Uniform{\Perm{Se}}{\second{S}}$}}
			\assert{$\{\LocalS(\invv{Trace,S}\land \size{S} = n \land p = \size{Trace}/2-n-1\land (\forall x.\ 0<x\le k \implies \Count{x,\second{S}}=m)\land$}\\
			\hspace*{0.4cm} \assert{$(\forall y.~ 0< y \le k\implies \size{s[i]} = \Count{i, \second{S}[1...p-1]} ))$\} } (Invariant for the last loop)\\
			$\whileR{p<\size{S}+1}{}$\\
			\hspace*{0.4cm} $\dassign{(e, i)}{S[p]};$\\
			\hspace*{0.4cm} $\record{(``read", ``S", p)}$\\
			\hspace*{0.4cm} $\dassign{s[i]}{s[i]+e};$\\
			\hspace*{0.4cm} $\record{(``write", ``s", i, \size{s[i]}+1)}$\\ 
			\hspace*{0.4cm} $\dassign{p}{p+1};$\\
			\assert{$\{\LocalS(\invv{Trace,S}\land \size{S} = n \land p = \size{Trace}/2-n-1\land (\forall x.\ 0<x\le k \implies \Count{x,\second{S}}=m)\land p> \size{S})\}$}\\
			\assert{$\{\LocalS(Trace = f(\second{S}) \land (f \text{ is bijective}))\}$}\\
			\centerline{End Const rule}
			\assert{$\{\LocalS(Trace = f(\second{S}) \land (f \text{ is bijective}))\land \Uniform{\Perm{Se}}{\second{S}}\}$}\\
			\assert{$\{\Uniform{f(Se)}{Trace}\}$}\\
			\centerline{End Unif-Idp rule}
			\assert{$\{\Uniform{f(Se)}{Trace}\sep \DD{SWO}\}$}
		\end{small}
	\end{minipage}
	\caption{Verification of Sampling Algorithm	\label{proofS}}
\end{figure*}

Let $A[i..j]$ denote the sub-array from $A[i]$ to $A[j]$ and $\Count{v,A}$ represent the number of occurrences of $v$ in the array $A$. Suppose $B$ is an array of tuples, then
let $\second{B}$ represent the array of all the second elements of the tuples in the array $B$.

Let $\inv{Trace}$ be a predicate on traces that holds
if and only if:

$Trace[1..3] = [ \text{oblishuffle}(n), (``read, ``D", 1), (``read",``D", 1)]\land$
\hspace*{0.4cm}$\forall x,\ 3<x\le\size{Trace}\implies$\\
\hspace*{\fill} $(x\%2 = 0 \implies Trace[x]=("write", "S", (x-2)/2))$
\hspace*{\fill} $\land(x\%2 = 1 \implies Trace[x]=("read", "S", (x-3)/2))$\\

Letting $\invv{Trace,S}$ be a predicate that holds
if and only if

$\inv{Trace[1..2n+3], S} \land Trace[2n+4] = \text{oblishuffle}(n) \land$\\
\hspace*{0.4cm}$\forall x, j.\ (2n+4<x\le\size{Trace} \land j = (x - (2n+3))/2)\implies$\\
\hspace*{\fill} $(x\%2 = 1 \implies Trace[x]=("read", "S", j)$\\
\hspace*{\fill} $\land(x\%2 = 0 \implies Trace[x]=("write", "s", \second{S}[j], \Count{\second{S}[j], \second{S}[1..j]})$\\

At the beginning, we have $\{\Local{n = m \times k}\}$.

Then, just before the first loop, we can apply random assignment rule, \textsc{Const} rule, random sample rule, and weak rule to get:\\
$\{\LocalS(Trace = [ \text{oblishuffle}(n), (``read, ``D", 1), (``read",``D", 1)] \land n = m \times k \land S = [] \land j = l = 1 \land e = e_{next} = D[1] \land D\in \Perm{D}\land SWO \in X(n,m))\}$ \\

Then, for the first loop, we use the loop invariant:\\
$\{ \LocalS(n = m \times k \land D\in \Perm{D}\land SWO \in X(n,m) \land l = \size{S} + 1 = (\size{Trace}-1)/2 = \Count{\true, SWO[1..k][1..j-1]}+1 \land\inv{Trace}\land(\forall x.\ 0<x\le k \implies \Count{\true, SWO[x][1..j-1]}=\Count{x,\second{S}})\land \size{Trace}\%2 = 1)\}$\\

For the inner loop, we use another loop invariant:\\
$\{ \LocalS(n = m \times k \land D\in \Perm{D}\land SWO \in X(n,m) \land l = \size{S} + 1 = (\size{Trace}-1)/2 = \Count{\true, SWO[1..k][1..j-1]}+\linebreak\Count{\true,SWO[1..i-1][j]}+1 \land\inv{Trace}\land\linebreak(\forall x.\ 0<x< i \implies \Count{\true, SWO[x][1..j]}=\Count{x,\second{S}})\land(\forall x.\ i\le x\le k \implies \Count{\true, SWO[x][1..j-1]}=\Count{x,\second{S}}))\}$\\

Both loop invariants can be proved by applying the random assignment rule, weak rule, and random if rule. After the two-layer loop, we know the first loop invariant holds and $l>n$. By
the weak rule, we obtain: \\
$\{\LocalS((\forall x.\ 0<x\le k \implies \Count{x,\second{S}}=m) \land \size{Trace} = 2n+3\land \inv{Trace} \land \size{S} = n)\}$\\
Note that
$\inv{Trace}$ is satisfied by at most one $Trace$ of a
particular length. So we know that $Trace$ has a
deterministic value at this point.

After the random choice following the loop, let $Se$ be an array containing $m$ occurrences of every number between 1 and $k$. Then we have that $\Uniform{\Perm{Se}}{\second{S}}$ in addition to the previous assertion, by the random sample rule. Then we apply the random assignment rule and weak rule several times to obtain the following assertion before the last loop:\\
$\{\LocalS(\invv{Trace,S}\land (\forall x.\ 0<x\le k \implies \Count{x,\second{S}}=m)\land \size{S} = n \land p = \size{Trace}/2-n-1\land (\forall y.~ 0< y \le k\implies \size{s[i]} = \Count{i, \second{S}[1...p-1]} ))\land \Uniform{\Perm{Se}}{\second{S}}\}$

Then we use the \textsc{Const} rule to add the Uniform assertion between the loop, and use the $\LocalS(\ldots)$ part of the above assertion as the last loop's invariant, which can be proved by applying the random loop, random if, random assignment, and weak rules. Finally, we obtain the above assertion and $p > \size{S}$. This information implies that the value of $Trace$ is a bijective function of $\second{S}$, which means $Trace$ also satisfies a uniform distribution. We can also apply the \textsc{Unif-Ind} rule to show $Trace$ is independent of $SWO$.
Note that since $Se$ is independent of the original database contents~$D$, we have trivially that $Trace$ is independent of~$D$.

\subsection{Path ORAM}\label{sec:pathORAM}
Path ORAM~\cite{pathORAM} is a (probabilistic) oblivious RAM algorithm. It allows a client to conceal its access pattern to
some remote storage. When
the client wants to read/write something from/to the remote storage, it calls
the function \textsc{Access}$(op, a, data*)$ where $op$ is the
type of access being performed
(read or write), $a$ is the \emph{virtual} (i.e.\ as seen
by the client) storage location being accessed,
and $data^*$ is either None (in the case of a read access)
or is a value to be written (in the case of a write access).
The \emph{physical} location of $a$ in the remote server is stored
in the global array $Q$ (originally called $\mathit{position}$ in~\cite{pathORAM}). This location
changes following the execution of the
algorithm, to hide (subsequent) access pattern of future
executions of the algorithm.

The original presentation of the algorithm included a
perfectly oblivious approximation~\cite[Figure 1]{pathORAM},
and bounded the statistical distance between it and the
practical version of the algorithm~\cite[Section 5]{pathORAM}.
Thus our focus is to prove the perfectly oblivious version
is indeed perfectly oblivious.

Our goal is to prove
that any two sequences of operations of the same length
produce indistinguishable memory access patterns, in the
sense that both are uniformly distributed over the same set
of possibilities.
A sequence of operations corresponds to a series of calls to
the \textsc{Access} function, e.g.\ 
a sequence
could be [(write,~$a$,~1), (read,~$a$,~None)], and would
correspond to two calls to the access function:
namely $\call{Access}{\text{write},a,1}$ followed by
$\call{Access}{\text{read},a,None}$. The distribution of
the memory access pattern
produced by this sequence should be identical to any other
sequence of operations of length~2.

The algorithm appears in \cref{fig:path} where, as earlier, we add ghost code to capture the observable memory access pattern and abstract from encryption and communication.
This figure follows the original~\cite[Figure 1]{pathORAM},
renaming the original $\mathit{position}$ array to~$Q$
for brevity.
We also add an additional
ghost variable $T'$ to record the initial value of $Trace$,
which will be useful
in the verification. The only two observable commands
are the function calls 
of $\WriteB{}$ and $\ReadB{}$, where their parameters $x$ and $l$ determine
the physical locations that are accessed.

\begin{figure}
	\begin{algorithmic}
		\Function{Access}{$op,a,data^*,\ghost{Trace}$}
		\State $\ghost{\dassign{T'}{Trace}}$
		\State $\dassign{x}{Q[a]}$
		\State $\rassign{Q[a]}{\{0..(2^L-1)\}}$
		\State $\dassign{l}{0}$
		\State $\whileR{l\le L}{}$
		\State \hspace*{0.4cm} $\dassign{S}{S\union \ReadB{P(x,l)}}$
		\State \hspace*{0.4cm} $\record{(``\ReadB{}", x, l)}$
		\State \hspace*{0.4cm} $\dassign{l}{l+1}$
		\State $\dassign{data}{\find{a, S}}$
		\State $\ifDs{op = \Write}{}$
		\State \hspace*{0.4cm} $\dassign{S}{(S - \{(a, data)\})\union \{(a, data^*)\}}$
		\State $\dassign{l}{L}$
		\State $\whileR{l\ge 0}{}$
		\State \hspace*{0.4cm} $\dassign{S'}{\{(a',data')\in S: P(x,l) = P(Q[a'],l)\}}$
		\State \hspace*{0.4cm} $\dassign{S'}{\select{Z, S'}}$
		\State \hspace*{0.4cm} $\dassign{S}{S-S'}$
		\State \hspace*{0.4cm} $\WriteB{P(x,l), S'}$
		\State \hspace*{0.4cm} $\record{(``\WriteB{}", x, l)}$
		\State \hspace*{0.4cm} $\dassign{l}{l-1}$
		\EndFunction
	\end{algorithmic}
	\caption{Rewritten path ORAM}
	\label{fig:path}
\end{figure}

We prove perfect obliviousness by proving that \textsc{Access}
maintains the following invariant on the observable memory
access trace~$Trace$: given any initial
memory access pattern~$Trace$ satisfying a fixed uniform distribution, the resulting access
pattern after calling \textsc{Access} (an extension of $Trace$)
still satisfies a
fixed uniform distribution.

We prove maintenance of a global invariant by proving
that the algorithm maintains its key implementation
invariant~\cite[Section 3]{pathORAM}, which we refer to in
this paper as the \emph{Main Invariant}:
each block is mapped to a uniformly random leaf
bucket in the tree (whose height is~$L$), i.e.\ every value in the
array $Q$ satisfies
the uniform distribution on $\{0..(2^L-1)\}$ and is independent of the
others. The Main Invariant is
encoded as an assertion in our logic using the $\Uniform{\cdot}{\cdot}$ and $*$ operators (see below).

Then the verification proceeds as follows.
Let $W = \{0...(2^L-1)\}$ and $n=\size{Q}$ which also means the number of (virtual) locations that can be accessed.
As shown in \cref{fig:path_proof}, we start with our desired invariant. It includes the Main Invariant which is simply
$\{(\Uniform{W}{Q[0]}\sep\Uniform{W}{Q[1]}\sep...\sep\Uniform{W}{Q[n]})\}$, plus the fact that
there exists a fixed set $Y$ where $Trace$ satisfies the uniform distribution on it, independently to the Main Invariant.

\begin{figure}
	\begin{algorithmic}
		\Function{Access}{$op,a,data^*,\ghost{Trace}$}\\
		$\assert{\{\text{Main Invariant} \sep \Uniform{Y}{Trace}\}}$
		\State $\ghost{\dassign{T'}{Trace}}$\\
		$\assert{\{\text{Main Invariant} \sep \Uniform{Y}{Trace}\land \Local{Trace = T'}\}}$
		\State $\dassign{x}{Q[a]}$\\
		$\assert{\{\Uniform{W}{Q[0]}\sep...\Uniform{W}{x}...\sep\Uniform{W}{Q[n]}\sep \Uniform{Y}{Trace}\land \Local{Trace = T'}\}}$
		\State $\rassign{Q[a]}{\{0..(2^L-1)\}}$\\
		$\assert{\{(\text{Main Invariant}\sep \Uniform{Y}{T'}\sep \Uniform{W}{x})\land \Local{Trace = T'}\}}$\\
		(Start \textsc{Const} rule)
		$\assert{\{\Local{Trace = T'}\}}$
		\State $\dassign{l}{0}$
		
		\State $\whileD{l\le L}{}$
		\State \hspace*{0.4cm} $\dassign{S}{S\union \ReadB{P(x,l)}}$
		\State \hspace*{0.4cm} $\record{(``\ReadB{}", x, l)}$
		\State \hspace*{0.4cm} $\dassign{l}{l+1}$
		\State $\dassign{data}{\find{a, S}}$
		\State $\ifDs{op = \Write}{}$
		\State \hspace*{0.4cm} $\dassign{S}{(S - \{(a, data)\})\union \{(a, data^*)\}}$
		\State $\dassign{l}{L}$
		\State $\whileD{l\ge 0}{}$
		\State \hspace*{0.4cm} $\dassign{S'}{\{(a',data')\in S: P(x,l) = P(Q[a'],l)\}}$
		\State \hspace*{0.4cm} $\dassign{S'}{\select{Z, S'}}$
		\State \hspace*{0.4cm} $\dassign{S}{S-S'}$
		\State \hspace*{0.4cm} $\WriteB{P(x,l), S'}$
		\State \hspace*{0.4cm} $\record{(``\WriteB{}", x, l)}$
		\State \hspace*{0.4cm} $\dassign{l}{l-1}$\\
		$\assert{\{\Local{Trace = T' \concat f(x)}\}}$
		(End \textsc{Const} rule)\\
		$\assert{\{(\text{Main Invariant}\sep \Uniform{Y}{T'}\sep \Uniform{W}{x})\land \Local{Trace = T' \concat f(x)}\}}$
		$\assert{\{\text{Main Invariant} \sep \Uniform{Y\times f(W)}{Trace}\}}$
		\EndFunction
	\end{algorithmic}
	\caption{Verification of path ORAM.}
	\label{fig:path_proof}
\end{figure}

The verification of the first three lines of code
is performed by unfolding the Main Invariant, and using the
random assignment rule, frame rule and the weak rule, then
refolding the Main Invariant.
Then we use the \textsc{Const} rule to carry all the information except $\Local{Trace = T'}$ to the end of this function as these facts are never modified. Thanks to using the \textsc{Const} rule,
reasoning proceeds via classical ($\Local{\cdot}$) reasoning.
It is easy to prove that at the end of the function we have $\Local{Trace = T' \concat f(x)}$, where $f(x) = [(\ReadB{},x,0),\ldots,(\ReadB{},x,L),\linebreak (\WriteB{},x,L),\ldots,(\WriteB{},x,0)]$. 

Finally, we convert this assertion using the third proposition in \cref{implication} and the proposition introduced above to prove the desired invariant.

\subsection{Path Oblivious Heap}\label{sec:pathOheap}

\newcommand{\Insert}{\textsc{Insert}\xspace}
\newcommand{\Add}{\textsc{Add}\xspace}
\newcommand{\Delete}{\textsc{Delete}\xspace}
\newcommand{\pos}{\mathit{pos}\xspace}
\newcommand{\ReadNRm}{\textsf{ReadNRm}\xspace}
\newcommand{\Evict}{\textsf{Evict}\xspace}
\newcommand{\UpdateMin}{\textsf{UpdateMin}\xspace}

\begin{figure*}
	\begin{tabular}{ll}
		\begin{minipage}{0.4\textwidth}
			\begin{algorithmic}
				\Function{Insert}{$k,v$}
				\State $\rassign{pos}{\{0,1,2,...,N-1\}}$
				\State $\call{Add}{B_{root}, (k,v,(pos,\tau))}$
				\State $\rassign{P}{\{0,1,2,...,N/2-1\}}$
				\State $\rassign{P'}{\{N/2,N/2+1,...,N-1\}}$
				
				\State $\call{Evict}{P}$
				\State $\call{UpdateMin}{P}$
				\State $\call{Evict}{P'}$
				\State $\call{UpdateMin}{P'}$
				
				\State \Return $(pos,\tau)$
				\EndFunction
			\end{algorithmic}
		\end{minipage} &
		\begin{minipage}{0.4\textwidth}
			\begin{algorithmic}
				\Function{Delete}{$pos,\tau$}
				\State $\call{ReadNRm}{pos,\tau}$
				\State $\call{Evict}{pos}$
				\State $\call{UpdateMin}{pos}$
				\EndFunction
			\end{algorithmic}
		\end{minipage}
	\end{tabular}
	\caption{The main interfaces of the Path Oblivious Heap.\label{fig:heap} See \ifExtended\cref{sec:subfunctions} \else supplementary Appendix C \fi for the sub-functions.}
\end{figure*}

\begin{figure*}
	Let $\inv{[pos_1,pos_2,\cdots ,pos_n]} = \Uniform{\{0..N-1\}}{pos_1} \sep \Uniform{\{0..N-1\}}{pos_2} \sep ..\sep \Uniform{\{0..N-1\}}{pos_n}$.
	
	\begin{algorithmic}
		\Function{Insert}{$k,v$}
		\State \assert{$\{\Uniform{T}{Trace} \sep \inv{E}\}$}
		\State $\rassign{pos}{\{0,1,2,...,N-1\}}$
		\State \assert{$\{\Uniform{T}{Trace} \sep \inv{E\concat [pos]}\}$}
		\State $\ghost{\assign{Trace'}{Trace}}$
		\State \assert{$\{\Uniform{T}{Trace'} \sep \inv{E\concat [pos]}\land \Local{Trace=Trace'}\}$}
		\State $\call{Add}{B_{root}, (k,v,(pos,\tau))}$
		\State \assert{$\{\Uniform{T}{Trace'} \sep \inv{E\concat [pos]}\land \Local{Trace=Trace'\concat TA(\text{root})}\}$}
		\State \assert{$\{\Uniform{T\times \{TA(\text{root})\}}{Trace} \sep \inv{E\concat [pos]}\}$}
		\State $\rassign{P}{\{0,1,2,...,N/2-1\}}$
		\State $\rassign{P'}{\{N/2,N/2+1,...,N-1\}}$
		\State \assert{$\{\Uniform{T\times \{TA(\text{root})\}}{Trace} \sep \inv{E\concat [pos]} \sep \Uniform{\{0..N/2-1\}}{P} \sep \Uniform{\{N/2..N-1\}}{P'}\}$}
		\State $\ghost{\assign{Trace'}{Trace}}$
		\State \assert{$\{\Uniform{T\times \{TA(\text{root})\}}{Trace'} \sep \inv{E\concat [pos]} \sep \Uniform{\{0..N/2-1\}}{P} \sep \Uniform{\{N/2..N-1\}}{P'} \land \Local{Trace=Trace'}\}$}
		\State $\call{Evict}{P}$
		\State $\call{UpdateMin}{P}$
		\State \assert{$\{\Uniform{T\times \{TA(\text{root})\}}{Trace'} \sep \inv{E\concat [pos]} \sep \Uniform{\{0..N/2-1\}}{P} \sep \Uniform{\{N/2..N-1\}}{P'} \land \Local{Trace=Trace' \concat  TE(P) \concat TU(P)}\}$}
		\State \assert{$\{\Uniform{T\times \{TA(\text{root})\} \times \{TE(p)\concat TU(p) ~|~ 0\le p<N/2\}}{Trace} \sep \inv{E\concat [pos]} \sep \Uniform{\{N/2..N-1\}}{P'} \}$}
		\State $\ghost{\assign{Trace'}{Trace}}$
		\State \assert{$\{\Uniform{T\times \{TA(\text{root})\} \times \{TE(p)\concat TU(p) ~|~ 0\le p<N/2\}}{Trace'} \sep \inv{E\concat [pos]} \sep \Uniform{\{N/2..N-1\}}{P'}\land \Local{Trace=Trace'} \}$}
		\State $\call{Evict}{P'}$
		\State $\call{UpdateMin}{P'}$
		\State \assert{$\{\Uniform{T\times \{TA(\text{root})\} \times \{TE(p)\concat TU(p) ~|~ 0\le p<N/2\}}{Trace'} \sep \inv{E\concat [pos]} \sep \Uniform{\{N/2..N-1\}}{P'}\land \Local{Trace=Trace' \concat  TE(P') \concat TU(P')} \}$}
		\State \assert{$\{\Uniform{T\times \{TA(\text{root})\} \times \{TE(p)\concat TU(p) ~|~ 0\le p<N/2\}\times \{TE(p)\concat TU(p) ~|~ N/2\le  p<N\}}{Trace'} \sep \inv{E\concat [pos]}\}$}
		\State \Return $(pos,\tau)$
		\EndFunction
	\end{algorithmic}
	
	\begin{algorithmic}
		\Function{Delete}{$pos,\tau$}
		\State \assert{$\{\Uniform{T}{Trace} \sep \inv{E} \land \Local{pos\in E}\}$}
		\State $\ghost{\assign{Trace'}{Trace}}$
		\State \assert{$\{\Uniform{T}{Trace'} \sep \inv{E} \land \Local{pos\in E} \land \Local{Trace=Trace'}\}$}
		\State $\call{ReadNRm}{pos,\tau}$
		\State \assert{$\{\Uniform{T}{Trace'} \sep \inv{E} \land \Local{pos\in E} \land \Local{Trace=Trace' \concat  TR(pos)}\}$}
		\State $\call{Evict}{pos}$
		\State \assert{$\{\Uniform{T}{Trace'} \sep \inv{E} \land \Local{pos\in E} \land \Local{Trace=Trace' \concat  TR(pos) \concat  TE(pos)}\}$}
		\State $\call{UpdateMin}{pos}$
		\State \assert{$\{\Uniform{T}{Trace'} \sep \inv{E} \land \Local{pos\in E} \land \Local{Trace=Trace' \concat  TR(pos) \concat  TE(pos) \concat  TU(pos)}\}$}
		\State (removing $pos_i$ from invariant)
		\State \assert{$\{\Uniform{T}{Trace'} \sep \inv{E\setminus [pos]} \sep \Uniform{\{0..N-1\}}{pos} \land \Local{pos\in E} \land \Local{Trace=Trace' \concat  TR(pos) \concat  TE(pos) \concat  TU(pos)}\}$}
		\State (using proposition 1.8)
		\State \assert{$\{\Uniform{T\times \{TR(p)\concat TE(p) \concat TU(p) ~|~ 0\le p < N\}}{Trace} \sep \inv{E\setminus [pos]}\}$}
		\EndFunction
	\end{algorithmic}
	
	\caption{Path Oblivious Heap (\cref{fig:heap}) Verification.
		\label{fig:heapV}}
\end{figure*}

The Path Oblivious Heap~\cite{pathOheap} provides the standard interfaces of a heap including element insertion,
deletion, finding the minimum, and extracting the minimum (this last being a combination of finding and deletion). The Path Oblivious Heap has been used to implement oblivious
sorting~\cite{pathOheap}, among other applications.
The Path Oblivious Heap is inspired by Path ORAM~\cite{pathORAM} and shares the same data tree structure and several sub-functions. However, unlike Path ORAM
which provides only one interface, the Path Oblivious Heap provides multiple
such which increases somewhat the complexity of its verification.

As a heap, the algorithm has two main interfaces. The \Insert function takes
two parameters: the key~$k$ and value~$v$ to be inserted; it returns
two values: $\pos$ and~$\tau$, the \emph{position} and \emph{timestamp}
of the inserted element that together uniquely identify the inserted element
in the heap. Timestamps are allocated deterministically: the first item
inserted into the heap has timestamp 0, the second 1, and so on. 
The \Delete function takes a position~$\pos$ and a
timestamp~$\tau$ and removes the corresponding element that they uniquely
identify.

This algorithm is designed to hide the contents of the key-value
parameters passed to \Insert (i.e.,\ to hide the heap data), but not
the kind of heap operations performed (i.e., whether \Insert or \Delete was called),
to an adversary who can
observe the algorithm's memory access pattern. Thus the adversary is assumed
to know what heap operations will be performed and in what order; although not
the parameters passed to those operations. As such, since timestamps are
deterministically allocated, they reveal no information and so, following~\cite{pathOheap}'s original presentation, for simplicity
we largely ignore
them henceforth.

Our verification applies to a perfectly oblivious approximation that
never fails; in practice the failure probability (and, thus, the statistical
distance between the real implementation and the perfectly oblivious
approximation that we verify) is bounded by a negligible quantity~\cite[Corollary 2]{pathOheap}, thereby
allowing \cref{lem:sd} to conclude obliviousness for the imperfect algorithm.

Our goal is to prove
that any two sequences of operations of the same length and that perform
the same types of operations in the same order
produce indistinguishable memory access patterns, in the
sense that both are uniformly distributed over the same set
of possibilities.
A sequence of operations corresponds to a series of calls to
the interfaces, e.g.\ 
a sequence
could be [\Insert, \Insert, \Delete{}]. The distribution of
the memory access pattern
produced by two executions of this sequence should be identical
regardless of the parameters passed in each.

\cref{fig:heap} depicts the oblivious heap algorithm~\cite[Section 3.3]{pathOheap}, where as earlier we add ghost code to capture the observable memory access pattern~$Trace$ and abstract from encryption and communication.
The various sub-functions (e.g., \Evict, \UpdateMin) also update $Trace$ to record
their memory-access patterns.

In \cref{fig:heapV}, we prove perfect obliviousness by proving that each interface
(1)~maintains an invariant $\inv{E}$, where $E$ is the sequence of existing elements' position in the heap, and (2) ensures that the resulting memory-access pattern is
uniformly distributed over a fixed set independent of the input parameters.
The invariant states that each position is independently uniformly distributed over
the fixed set of possible positions:
\[
\inv{[pos_1,pos_2,\cdots ,pos_n]} = \Uniform{\{0..N-1\}}{pos_1} \sep \Uniform{\{0..N-1\}}{pos_2} \sep ..\sep \Uniform{\{0..N-1\}}{pos_n}
\]
The preconditions in~\cref{fig:heapV} should be read as quantifying
over~$T$, the set over which the historical memory-access pattern is uniformly
distributed.

We put the classical, Hoare logic
verification of the sub-functions and the interface of finding minimum in the next section (they are totally deterministic).
The classical specification for each sub-function
states that it adds to the $Trace$ a fixed
access pattern that depends only on the input parameter. For instance, \Evict's
classical Hoare logic specification is:\\

\noindent$\triple{Trace = Trace'}{\Evict(x)}{Trace = Trace' \concat TE(x)}$,
where~$TE(x)$ (``Trace of Evict'')
abbreviates the access pattern for \Evict for input
parameter~$x$.\\ 

We follow this same naming convention throughout: e.g. $TU(x)$ is the
memory-access pattern of $\UpdateMin(x)$. For a set~$X$, we write $TE(X)$ to mean
$\{TE(x)\ |\ x\in X\}$
and so on.

\Delete's precondition makes sure its input is valid (we only delete existing elements).
We first add a ghost variable $Trace'$ to record the initial value of 
$Trace$ and then use $\Local{\cdot}$ (certain) reasoning for the following three function calls. Finally we convert the assertion to the desired one by
\cref{implication}. To do so, we use the fact that
$TR, TE, TU$ are bijective and always produce the sequence with the same length on their valid inputs (from 0 to $N-1$) respectively, to satisfy the assumption of the 8th proposition of \cref{implication}. 

Finally, the verification of \Insert shares the same idea as \Delete, repeated several times. Note, $TA(\text{root})$ is a fixed sequence because sub-function \Add's memory-access pattern is independent of \Add's arguments. $\pos$ is the position of the inserted element and is never used or leaked in this function. It is recorded in the invariant and will be released (and leaked) when it is deleted; however, as explained in
\cref{sec:motivation}, doing so reveals nothing since $\pos$ was chosen uniformly
and independently of secrets.

\section{Path Oblivious Heap Deterministic Sub-functions}\label{sec:subfunctions}

The path oblivious heap is implemented internally by two arrays: there is an array $tree$ of tree nodes that represents the tree, and an array $min$ where $min[i]$ stores the minimum element of the sub-tree rooted at node $tree[i]$.
The length of both arrays is $2N$; however index 0 is unused, as is
standard for using an array to represent a heap.

The reads and writes to both arrays are observable to the attacker. 

Each node in the tree has a fixed number of positions for storing elements, where an element has the form $(k, v, \tau)$ for key~$k$, value~$v$
and timestamp~$\tau$.
The number of positions for the root node ($n_\text{root}$) is decided by the algorithm's security parameter~\cite{pathOheap}, whereas the number of positions for the other nodes is 4. All of them store a dummy element~\text{DUMMY} initially.

We use ghost code to record the accesses to the $tree$ and $min$ arrays. When we access a position $tree[i][j]$, we add $("read", i, j)$ or $("write", i , j)$ into the memory-access trace depending on which type of access was performed. Similarly, when we access a position $min[i]$, we add $("readMin", i)$ or $("writeMin", i)$ into the trace. When we read an entire node, we use $("readAll", i)$ to denote $[("read", i, 0), ("read", i, 1)], ... $ up to the number of positions for the node; we use corresponding notation for write accesses.

Given an integer $p$ such that $0\le p< N$, we define $\call{path}{p}$ as the sequence of indexes of the path from the root to the $p$th leaf of the tree. For example, $\call{path}{0} = [1,2,4,...,N]$. We also define $\call{pathR}{p}$ as the reversed sequence of $\call{path}{p}$.

Then the sub-functions and corresponding specifications are in \cref{fig:subf}. The corresponding functions that define their memory-access patterns are in \cref{fig:subf2}.

\begin{figure*}
	\[
	\call{TA}{x}= \call{TD}{x} =
	\begin{cases}
		[("read", 1, 0), ("write", 1, 0), ..., ("read", 1, n_\text{root}), ("write", 1, n_\text{root})],& \text{if }x = 1 \text{ (root)}\\
		[("read", x, 0), ("write", x, 0), ... , ("read", x, 3), ("write", x, 3)],              & \text{otherwise}
	\end{cases}
	\]
	
	\ \\
	
	$\call{TR}{x} = \call{TD}{\call{path}{p}[0]} \concat \call{TD}{\call{path}{p}[1]} \concat\cdots\concat\call{TD}{\call{path}{p}[\call{log}{N}]}$
	
	\ \\
	
	$\call{TU}{x} = [("readAll",\call{pathR}{x}[0]), ("writeMin", \call{pathR}{x}[0])]\concat [("readAll",\call{pathR}{x}[1]), ("readMin", 2*\call{pathR}{x}[1]), ("readMin", 2*\call{pathR}{x}[1]+1), ("writeMin", \call{pathR}{x}[1])]\concat [("readAll",\call{pathR}{x}[2]), ("readMin", 2*\call{pathR}{x}[2]), ("readMin", 2*\call{pathR}{x}[2]+1), ("writeMin", \call{pathR}{x}[2])]\concat \cdots \concat [("readAll",\call{pathR}{x}[\call{log}{N}]), ("readMin", 2*\call{pathR}{x}[\call{log}{N}]), ("readMin", 2*\call{pathR}{x}[\call{log}{N}]+1), ("writeMin", \call{pathR}{x}[\call{log}{N}])]$
	
	\ \\
	
	$\call{TE}{x} = [("readAll",\call{path}{x}[0]), ("readAll",\call{path}{x}[1]), \cdots, ("readAll",\call{path}{x}[\call{log}{N}])] \concat [("writeAll",\call{path}{x}[0]), ("writeAll",\call{path}{x}[1]), \cdots, ("writeAll",\call{path}{x}[\call{log}{N}])]$
	\caption{Path Oblivious Heap's trace functions \label{fig:subf2}}
\end{figure*}

\begin{figure*}
	\begin{tabular}{ll}
		\begin{minipage}{0.5\textwidth}
			\begin{algorithmic}
				\Function{add}{$i, k, v, \tau, Trace$}
				\State \assert{$\{Trace = Trace'\}$}
				\State $\dassign{s}{\false}$
				\State $\dassign{j}{0}$
				\State $\whileD{j < \call{length}{tree[i]}}{}$
				\State \hspace*{0.4cm} $\dassign{w}{tree[i][j]}$
				\State \hspace*{0.4cm} $\record{("read", i, j)}$
				\State \hspace*{0.4cm} $\ifDs{w = \text{DUMMY} \land (!s)}{}$
				\State \hspace*{0.8cm} $\dassign{w}{(k,v,\tau)}$
				\State \hspace*{0.8cm} $\dassign{s}{\true}$
				\State \hspace*{0.4cm} $\dassign{tree[i][j]}{w}$
				\State \hspace*{0.4cm} $\record{("write", i, j)}$
				\State \hspace*{0.4cm} $\dassign{j}{j+1}$
				\State \assert{$\{Trace = Trace' + TA(i)\}$}
				\EndFunction
			\end{algorithmic}
			\begin{algorithmic}
				\Function{del}{$i, \tau, Trace$}
				\State \assert{$\{Trace = Trace'\}$}
				\State $\dassign{j}{0}$
				\State $\whileD{j < \call{length}{tree[i]}}{}$
				\State \hspace*{0.4cm} $\dassign{w}{tree[i][j]}$
				\State \hspace*{0.4cm} $\record{("read", i, j)}$
				\State \hspace*{0.4cm} $\ifDs{\tau = w[2]}{}$
				\State \hspace*{0.8cm} $\dassign{w}{\text{DUMMY}}$
				\State \hspace*{0.4cm} $\dassign{tree[i][j]}{w}$
				\State \hspace*{0.4cm} $\record{("write", i, j)}$
				\State \hspace*{0.4cm} $\dassign{j}{j+1}$
				\State \assert{$\{Trace = Trace' + TD(i)\}$}
				\EndFunction
			\end{algorithmic}
		\end{minipage}
		&
		\begin{minipage}{0.5\textwidth}
			\begin{algorithmic}
				\Function{readNRm}{$p, \tau, Trace$}
				\State \assert{$\{Trace = Trace'\}$}
				\State $\dassign{j}{0}$
				\State $\whileD{j < \call{log}{2N}}{}$
				\State \hspace*{0.4cm} $\textsc{del}({\call{path}{p}[j],\tau,Trace})$
				\State \hspace*{0.4cm} $\dassign{j}{j+1}$
				\State \assert{$\{Trace = Trace' + TR(p)\}$}
				\EndFunction
			\end{algorithmic}
		\ \\
			\begin{algorithmic}
				\Function{updateMin}{$p, Trace$}
				\State \assert{$\{Trace = Trace'\}$}
				\State $\dassign{i}{0}$
				\State $\whileD{i < \call{log}{2N}}{}$
				\State \hspace*{0.4cm} $\dassign{j}{\call{pathR}{p}[i]}$
				\State \hspace*{0.4cm} $\dassign{A}{tree[j]}$
				\State \hspace*{0.4cm} $\record{("readAll", j)}$
				\State \hspace*{0.4cm} $\ifDs{i>0}{}$
				\State \hspace*{0.8cm} $\dassign{A}{A+min[2j]+min[2j+1]}$
				\State \hspace*{0.8cm} $\record{("readMin", 2j)}$
				\State \hspace*{0.8cm} $\record{("readMin", 2j+1)}$
				\State \hspace*{0.4cm} $\dassign{min[j]}{\call{min}{A}}$
				\State \hspace*{0.4cm} $\record{("writeMin", j)}$
				\State \hspace*{0.4cm} $\dassign{i}{i+1}$
				\State \assert{$\{Trace = Trace' + TU(p)\}$}
				\EndFunction
			\end{algorithmic}
		\end{minipage}
	\end{tabular}\\\ \\

	\begin{algorithmic}
		\Function{Evict}{$p, Trace$}
		\State \assert{$\{Trace = Trace'\}$}
		\State $\dassign{i}{0}$
		\State $\dassign{A}{[]}$
		\State $\whileD{i < \call{log}{2N}}{}$
		\State \hspace*{0.4cm} $\dassign{A}{A + tree[\call{path}{p}[i]]}$
		\State \hspace*{0.4cm} $\record{("readAll", \call{path}{p}[i])}$
		\State \hspace*{0.4cm} $\dassign{i}{i+1}$
		\State $\dassign{A}{\call{Evict\_Locally}{A}}$
		\State $\whileD{i < \call{log}{2N}}{}$
		\State \hspace*{0.4cm} $\dassign{tree[\call{path}{p}[i]]}{A[i]}$
		\State \hspace*{0.4cm} $\record{("writeAll", \call{path}{p}[i])}$
		\State \hspace*{0.4cm} $\dassign{i}{i+1}$
		\State \assert{$\{Trace = Trace' + TE(p)\}$}
		\EndFunction
	\end{algorithmic}

	\caption{Path Oblivious Heap's deterministic sub-functions\label{fig:subf}. \textsf{Evict\_Locally} moves elements from the root towards the leaf on a given path, as described in \cite[Section 3.2]{pathOheap}. It operates over the private memory~$A$ and so its memory accesses are unobservable to the attacker.}
\end{figure*}

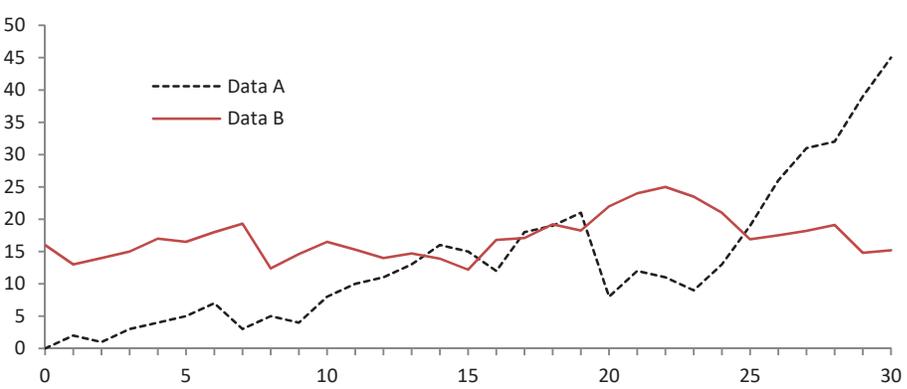
\end{document}
\endinput
